# Progress Towards the Calculation of Time-Dependent, Viscous, Compressible Flows using Active Flux Scheme

Oliviu Şugar-Gabor

*Aeronautical and Mechanical Engineering Group*

*University of Salford*

*Salford, M5 4WT, UK*

o.sugar-gabor@salford.ac.uk

## ABSTRACT

The Active Flux scheme is a Finite Volume scheme with additional degrees of freedom. It makes use of a continuous reconstruction and does not require a Riemann solver. An evolution operator is used for the additional degrees of freedom on the cell boundaries. This paper presents progress towards the computation of one-dimensional, viscous, compressible flows using Active Flux scheme. An evolution operator for both linear and nonlinear hyperbolic conservation systems is presented and then a novel extension is made to include source terms. Applications are made on the Euler equations and a hyperbolic formulation of the diffusion equation. Lastly, for the compressible Navier-Stokes equations, a hyperbolic formulation is presented together with a novel operator splitting approach. These allow for the Active Flux evolution operators to be applied to the numerical computation of viscous, compressible flows.

## 1. INTRODUCTION

The Active Flux (AF) scheme, originally introduced by Eymann and Roe in a series of works (Eymann and Roe, 2011a, 2011b, 2013) is a Finite Volume (FV) scheme with additional degrees of freedom on the cell boundaries. These additional nodes allow for third-order accuracy to be achieved, though extensions to higher orders are also possible. What distinguishes AF from other FV scheme are three aspects. First, variables have $C^0$ continuity across cell boundaries, even for nonlinear problems where discontinuities may be present, thus not requiring a Riemann solver. The $C^0$ continuity also reduces the number of independent degrees of freedom in each cell compared to other high-order methods such as Discontinuous Galerkin, thus reducing the memory requirements. Second, the variable values at the additional degrees of freedom on the cell boundaries are evolved independently from the cell-average values, using (non-conservative) compact-stencil schemes most suited to the mathematical and physical properties of the equation(s) being solved. Third, the flux across cell boundaries is calculated by numerically integrating the physical flux using quadrature.

The calculation of nonlinear conservation laws has been a major area of research over the past half-century. Suitable numerical methods must be able to capture discontinuities (shock waves in high-speed fluid flows), preserve fundamental physical quantities (vorticity, mass) as well as avoid non-physical solutions (expansion shocks). Typical FV schemes make use of dimensional splitting, replacing the multi-dimensional problem by one-dimensional problems through each cell boundary. The approach requires significant grid refinement to capture multi-dimensional features. LeVeque (2002) provides an excellent overview and introduction to FV methods.

Recently published research (see Barsukow et al., 2019 and Barsukow, 2021) indicates that the AF scheme is a very promising candidate to providing relatively simple, vorticity-preserving, shock-capturing and true multi-dimensional solutions to nonlinear conservation laws. For linear equations such as simple advection or linearized acoustics, the AF scheme has been studied in detail. The works of Eymann and Roe (2011a, 2011b, 2013) and Fan (2017) cover these items. For nonlinear equations, approximate evolution operators for Burgers' equation (Eymann and Roe, 2013, Maeng, 2017) and the Euler equations (Maeng, 2017, Barsukow, 2021) have been studied. More recently, He (2021) has made significant progress towards the application of AF schemes to the Navier-Stokes equations for laminar, subsonic flows, using a complicated evolution operator based on the Cauchy-Kovalevskaya / Lax-Wendroff procedure and an evaluation of higher order spatial derivatives using an approach inspired from the method of spherical means.

In the present work, further progress is presented towards the application of the AF scheme to compressible viscous flows with shock waves and more complex flow behaviour. Section 2 provides an overview of the AF scheme and a first simple example application on the Burgers equation. Section 3 details the nonlinear hyperbolic system evolution operator first derived by Barsukow (2021) and its application to the one-dimensional Euler equations. Section 4 presents a novel extension of the evolution operator to hyperbolic systems including source terms, as well as an application of the AF scheme to a hyperbolic formulation of the diffusion equation. Finally, Section 5 details the hyperbolic formulation of the Navier-Stokes equations derived by Nishikawa (2011), and presents, to the author's best knowledge, the first derivation and application of the AF scheme to compressible viscous flows including shock waves. Such flows are of special interest to aeronautical engineers due to their relatively common occurrence in transonic and supersonic flight. The paper ends with conclusions and a discussion of future work.

## 2. OVERVIEW OF THE ACTIVE FLUX SCHEME

### 2.1. Finite volume discretisation

To provide a brief but intuitive description of the AF scheme, consider the nonlinear one-dimensional scalar conservation law:

$$\frac{\partial U}{\partial t}+\frac{\partial F}{\partial x}=0 \tag{1}$$

where $U(t,x) : \mathbb{R}^{+} \times \mathbb{R} \rightarrow \mathbb{R}$ and $F\big(U(t,x)\big)$ is the flux function.

Following the approach used in FV methods, the space domain is divided into $N$ non-overlapping computational cells $\Omega_i$ , $i = 1, N$, each cell being centred around grid point $x_i$ and having a left $x_{i-1/2}$ and a right $x_{i+1/2}$ boundary. The time domain is divided into $M$ timesteps $t^n$ , $n = 1, M$. The grid spacing $\Delta x = x_{i+1/2} - x_{i-1/2}$ and time step $\Delta t = t^{n+1} - t^n$ are taken to be constant, though extensions to variable stepping do not pose any significant problems.

Define cell-average values:

$$\bar{U}_i(t)=\frac{1}{\Delta x}\int_{x_{i-1/2}}^{x_{i+1/2}} U(t,x)dx \tag{2}$$

By integrating (1) over cell $\Omega_i$ and in time between $t^n$ and $t^{n+1}$, the classical discretization of a conservation law typical to FV methods is obtained:

$$\bar{U}_i^{n+1}=\bar{U}_i^{n}-\frac{\Delta t}{\Delta x}\Big(\bar{F}\big(x_{i+1/2}\big)-\bar{F}\big(x_{i-1/2}\big)\Big) \tag{3}$$

where $\bar{U}_i^n$ represents the cell-average value in cell $\Omega_i$ at time $t^n$. As in many FV schemes, this value is associated with grid point $x_i$. $\bar{F}\big(x_{i+1/2}\big)$ is the flux through the cell boundary $x_{i+1/2}$ and is given by:

$$\bar{F}\big(x_{i+1/2}\big)=\frac{1}{\Delta t}\int_{t^n}^{t^{n+1}} F\Big(U\big(t,x_{i+1/2}\big)\Big)dt \tag{4}$$

### 2.2. Solution reconstruction

In the AF method, the boundary points are used together with the cell-centre point to reconstruct the solution inside each cell. The cell average is retained as one of the degrees of freedom. Furthermore, the reconstruction is $C^0$ continuous everywhere in the computational domain.

In one-dimension, a parabola is the natural choice for achieving third-order accuracy:

$$U_{recon}(x) = -3(2\overline{U}_i - U_{i-1/2} - U_{i+1/2})\frac{(x - x_i)^2}{\Delta x^2} + (-U_{i-1/2} + U_{i+1/2})\frac{x - x_i}{\Delta x} + \frac{1}{4}(6\overline{U}_i - U_{i-1/2} - U_{i+1/2}), x \in [x_{i-1/2}, x_{i+1/2}] \quad (5)$$

Here, $U_{i-1/2}$ and $U_{i+1/2}$ are the variable values for the boundary points. Dependency on time has not been included in (5) to enhance clarity of notation, but it is implicitly understood.

Details about two-dimensional reconstructions on unstructured grids can be found in Maeng (2017), and on structured grids in Barsukow et al. (2019). He (2021) provides some suitable reconstructions for increasing the order to accuracy above third order, though it must be noted that suitable modifications must also be made in other sections of the scheme.

### 2.3. Evolution operator

The key stage of the method is the evolution of the point degrees of freedom on the cell boundaries. The reconstruction using data from $t^n$ plays the role of an initial condition. The evolution stage advances $U_{i-1/2}$ and $U_{i+1/2}$ from $t^n$ to $t^{n+1}$ using a scheme most suited to the mathematical and physical properties of the equation(s) being solved.

To evolve the point values, consider the non-conservative form of (1):

$$\frac{\partial U}{\partial t} + A(U)\frac{\partial U}{\partial x} = 0 \quad (6)$$

where $A(U) = \partial F/\partial U$.

For simplicity, introduce the notation:

$$U_{recon}(t^n, x) = U_0(x) \quad (7)$$

The updated values at the boundary points $U(t^{n+1}, x_{i+1/2})$ are determined based on the theory of characteristics and are given by:

$$U(t^{n+1}, x_{i+1/2}) = U_0(x_0) \quad (8)$$

Here, $x_0$ represents the foot of the characteristic traced back from the boundary point $x_{i+1/2}$ at time $t^{n+1}$.

The characteristics are straight lines along which the solution is constant. It must be noted however that for a nonlinear equation, the characteristics may intersect, generating discontinuities even from smooth initial data. This is due to the fact that the characteristic speed $A(U)$ is function of the solution itself.

The coordinate of the characteristic foot can be determined by:

$$x_0 = x_{i+1/2} - A(U_0(x_0))\Delta t \quad (9)$$

Barsukow (2021) shows that a relatively simple fixed-point iteration procedure can be used to solve (9):

$$x_0{}^{(0)} = x_{i+1/2}$$
$$x_0{}^{(k+1)} = x_{i+1/2} - A\left(U_0\left(x_0{}^{(k)}\right)\right)\Delta t \tag{10}$$

The procedure requires $(k)$ iterations to solve for the position of the characteristic foot $x_0$ to the $k^{th}$ order of accuracy.

The evolution of the point degrees of freedom on the cell boundaries is done twice, once for the full timestep $U(t^{n+1}, x_{i+1/2})$ and once for the half timestep $U(t^{n+1/2}, x_{i+1/2})$. Next, Simpson's 1/3 quadrature rule is used for the flux integral:

$$\int_{t^n}^{t^{n+1}} F\left(U(t, x_{i+1/2})\right) dt \cong \frac{\Delta t}{6}\left(F\left(U(t^n, x_{i+1/2})\right) + 4F\left(U(t^{n+1/2}, x_{i+1/2})\right) + F\left(U(t^{n+1}, x_{i+1/2})\right)\right) \tag{11}$$

In the formulation adopted in this work, the AF scheme is explicit in time, though an implicit formulation for linear equations has been published (Nishikawa and Roe, 2016). As an explicit scheme, there is a CFL condition that must be satisfied:

$$\Delta t < \frac{\Delta x}{A_{max}} \tag{12}$$

where $A_{max}$ is the maximum value of the characteristic speed.

In practice, the condition is implemented with the help of the widely used Courant number $\sigma$:

$$\Delta t = \sigma \frac{\Delta x}{A_{max}} \tag{13}$$

### 2.4. General overview of the scheme

All the steps of the AF scheme have now been fully defined and can be summarised as follows:

Step 1: Given cell averages $\bar{U}_i^n$ and additional degrees of freedom $U_{i-1/2}^n$, $U_{i+1/2}^n$, calculate reconstruction coefficients for each cell using (5).

Step 2: Loop over all cell boundaries. For each cell boundary, determine the position of the characteristic foot $x_0$ using (10), once for the full timestep $\Delta t$ and once for the half timestep $\Delta t/2$. Update $U(t^{n+1}, x_{i+1/2})$ and $U(t^{n+1/2}, x_{i+1/2})$ using (8).

Step 3: Loop over all cell boundaries and determine the flux using the quadrature (11).

Step 4: Update the cell averages using the conservative update (3).

### 2.5. Application to Burgers equation

A simple application of the AF scheme to the Burgers equation is considered as an initial example.

$$\frac{\partial U}{\partial t} + \frac{\partial U^2/2}{\partial x} = 0 \tag{14}$$

Let $x \in [0, 1]$ and the initial condition be:

$$U_0(x) = \frac{1}{2\pi}\sin(2\pi x) \tag{15}$$

While the initial condition is a smooth sine function, the solution will evolve towards a shock located at $x = 0.5$ after a time of minimum $t = 1s$. Figure 1 shows the obtained results and the AF scheme's ability to capture shock formation from smooth initial data.

Of equal importance is checking the AF scheme's order of convergence. A run with $N = 2560$ cells is taken as a substitute "exact" solution. The $L2$ norm of the error between the solution at increasing grid density and the substitute "exact" solution is calculated at $t = 0.5s$ and plotted in Figure 2. The coarsest grid level has $N = 5$ cells and the number of cells is doubled for each subsequent grid level. It can be observed that the AF scheme converges at third-order accuracy for the Burgers equation.

## 3. NONLINEAR HYPERBOLIC SYSTEMS AND THE EULER EQUATIONS

### 3.1. The Euler equations

Consider the one-dimensional system of conservation laws:

$$\begin{gathered} \frac{\partial \boldsymbol{U}}{\partial t} + \frac{\partial \boldsymbol{F}}{\partial x} = \boldsymbol{0} \\ \boldsymbol{U}(t_0, x) = \boldsymbol{U}_0(x) \end{gathered} \tag{16}$$

where $\boldsymbol{U}(t,x) : \mathbb{R}^+ \times \mathbb{R} \to \mathbb{R}^m$, $\boldsymbol{F}\big(\boldsymbol{U}(t,x)\big)$ is the flux and $\boldsymbol{U}_0(x)$ is the initial condition.

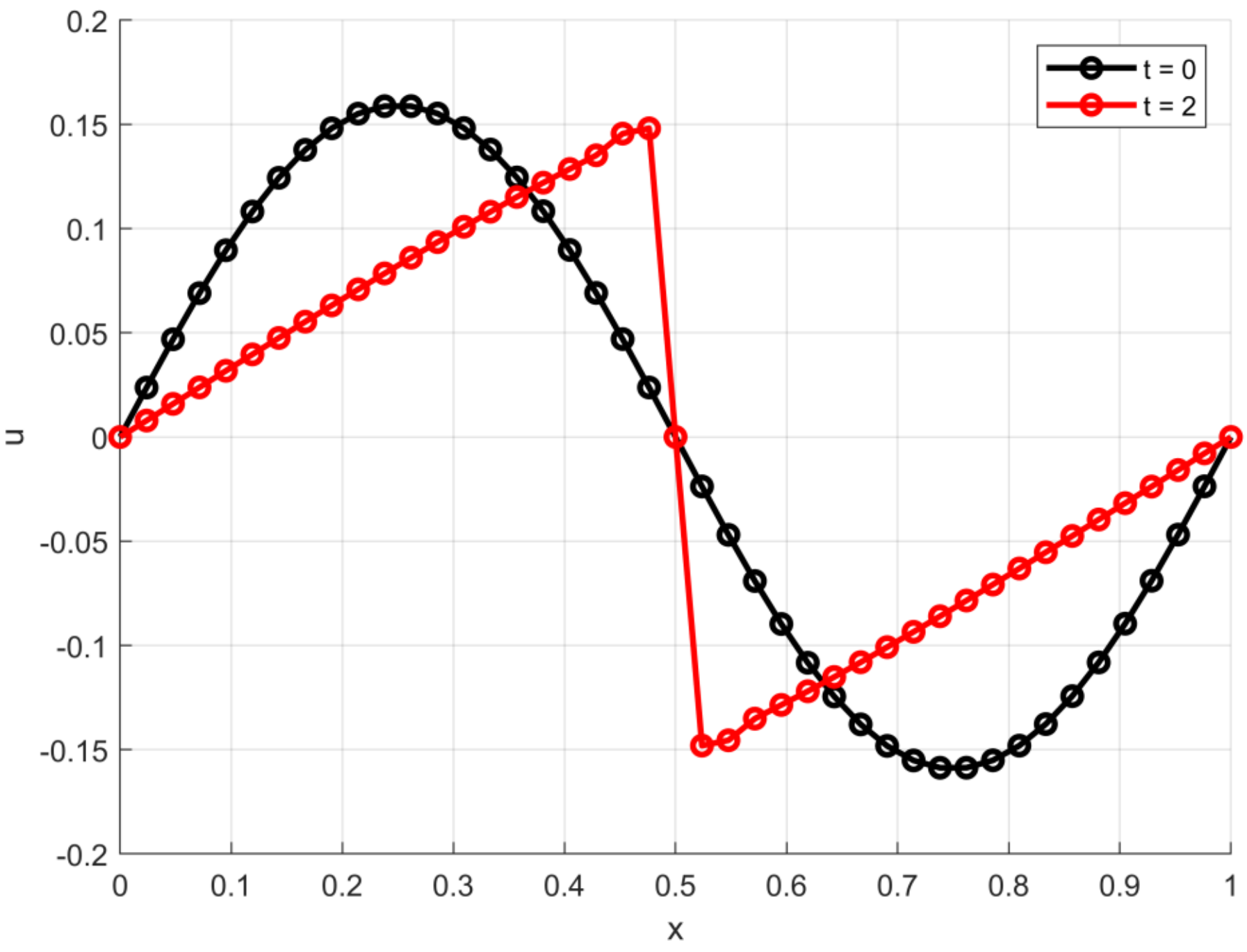

**Figure 1. Shock formation from smooth initial data using Burgers' equation. Solution was obtained at $t = 2\ s$ with $N = 21$ grid cells and $\sigma = 0.7$. Plotted results include both boundary point values and cell-average values.**

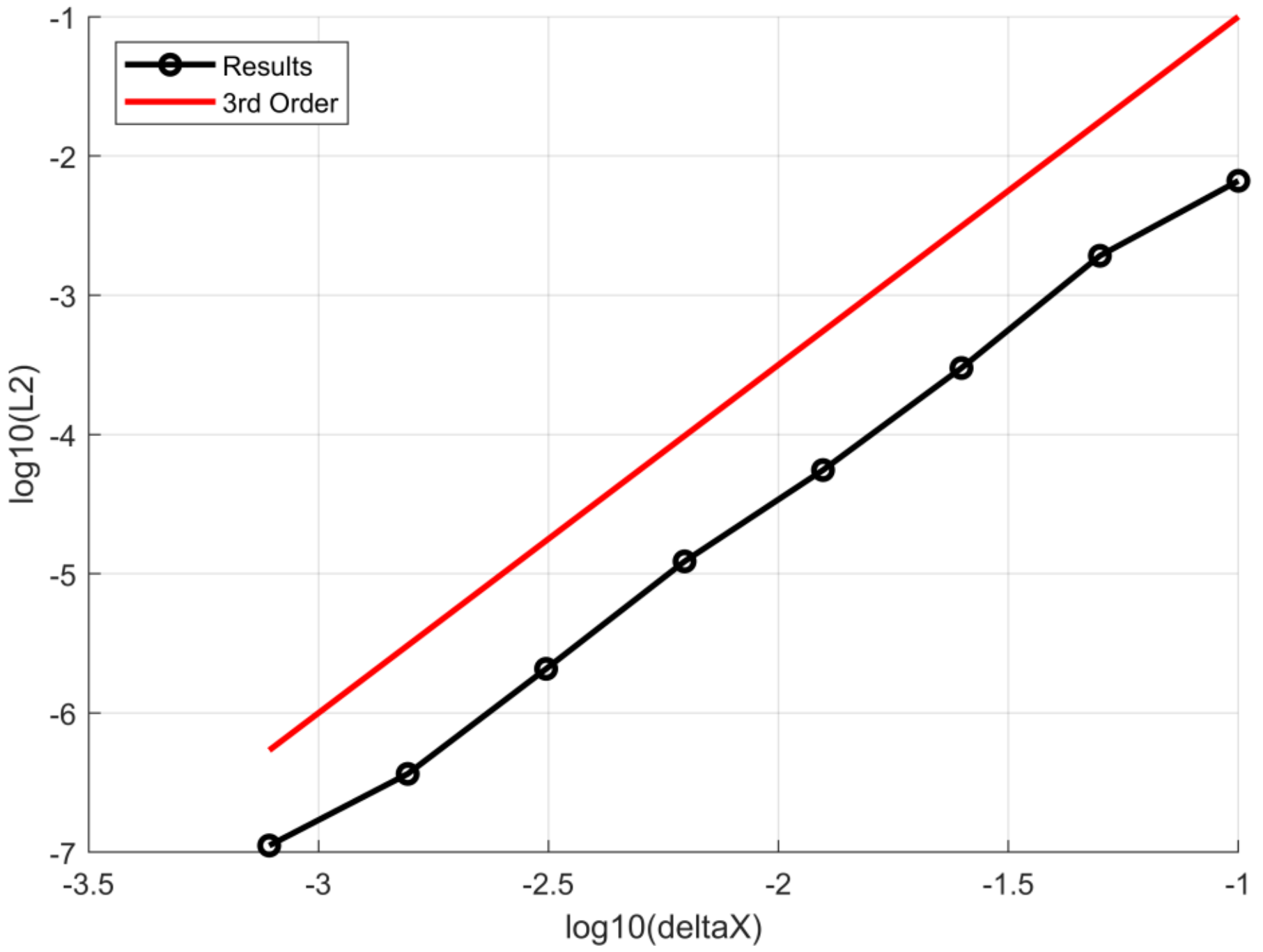

**Figure 2. Order of convergence of the solution to Burgers' equation at $t = 0.5s$ as measured by the $L2$ error with respect to an extremely fine grid. A line corresponding to 3rd order is shown for reference.**

The system is taken to be hyperbolic, that is the Jacobian matrix $\boldsymbol{A} = \partial \boldsymbol{F} / \partial \boldsymbol{U}$ has only real eigenvalues and has a complete set of linearly independent eigenvectors. The Euler equations describing the compressible, adiabatic flow of an inviscid fluid are a well-known example, for which the various terms appearing in (16) are:

$$\boldsymbol{U} = \begin{bmatrix} \rho \\ \rho u \\ \rho E \end{bmatrix} \quad \boldsymbol{F} = \begin{bmatrix} \rho u \\ \rho u^2 + p \\ \rho u H \end{bmatrix} \tag{17}$$

Here, $\rho$ is the density, $u$ is the flow velocity, $E$ is the specific total energy, $p$ is the pressure and $H = E + p/\rho$ is specific total enthalpy. It is assumed the fluid also satisfies the equation of state for ideal gases, which can be written in a convenient form as $p = (\gamma - 1)(\rho E - \rho u^2/2)$, with $\gamma$ being the ratio of specific heats.

Results for the one-dimensional Euler equations using the AF scheme have been published by Fan (2017) or He (2021). The approach used was based on an operator splitting approach, in which the nonlinear advection is solved first, using a method based on approximate characteristic tracing, followed by solving for the acoustic (pressure) component, using the method of spherical means. A simpler predictor-corrector approach was developed by Barsukow (2021), based on the mathematical properties of hyperbolic systems of equations. The approach used in this work follows that proposed by Barsukow (2021).

### 3.2. Evolution operator

Let $\boldsymbol{L}$ be the matrix of left eigenvectors, $\boldsymbol{R}$ be the matrix of right eigenvectors and $\boldsymbol{\Lambda}$ be the diagonal eigenvalues matrix, $\boldsymbol{\Lambda} = \mathrm{diag}(\lambda_1, \lambda_2, \ldots)$. It is also known that $\boldsymbol{R}^{-1} = \boldsymbol{L}$.

Details on the eigenvalues and eigenvectors of the Euler equation can be found in many textbooks, see for example Anderson (1995). By multiplying (16) to the left by $\boldsymbol{L}$ and defining the characteristic variables in the general sense as $\partial \mathbf{W} = \boldsymbol{L}\partial \boldsymbol{U}$, the characteristic system is obtained:

$$\frac{\partial \mathbf{W}}{\partial t} + \boldsymbol{\Lambda}\frac{\partial \mathbf{W}}{\partial x} = \mathbf{0} \tag{18}$$

The one-dimensional Euler equations are one of the cases where the characteristic equations and variables are decoupled. For higher dimensions however, this is no longer true. Nonetheless, the variation $\partial \mathbf{W}$ can always be defined and can be used as basis to inspire a suitable evolution operator for the AF scheme.

For a system of $m$ equations, each characteristic variable $\partial W_k\,, k = 1, m$ is propagated with a speed $\lambda_k$ along the characteristic curve $C^k$. Considering two instances in time, $t^n$ and $t^{n+1} = t^n + \Delta t$, the following equation holds along the corresponding characteristic curve:

$$\partial W_k(t^{n+1}, x) = \partial W_k(t^n, x - \lambda_k \Delta t)\,, k = 1, m \tag{19}$$

Following from (19) and remembering that $\partial \mathbf{W} = \boldsymbol{L}\partial \boldsymbol{U}$:

$$\begin{aligned} \partial \boldsymbol{U}(t^{n+1}, x) &= \boldsymbol{L}^{-1}\,\partial \mathbf{W}(t^{n+1}, x) = \boldsymbol{R}\,\partial \mathbf{W}(t^{n+1}, x) \\ &= \sum_{k=1}^{m} \boldsymbol{R}_k\,\partial W_k(t^{n+1}, x) = \sum_{k=1}^{m} \boldsymbol{R}_k\,\partial W_k(t^n, x - \lambda_k \Delta t) \\ &= \sum_{k=1}^{n} \boldsymbol{R}_k\,\boldsymbol{L}_k\,\partial \boldsymbol{U}(t^n, x - \lambda_k \Delta t) \end{aligned} \tag{20}$$

where $\boldsymbol{R}_k$ is the $k$ right eigenvector (column $k$ of the $\boldsymbol{R}$ matrix) and $\boldsymbol{L}_k$ is the $k$ left eigenvector (row $k$ of the $\boldsymbol{L}$ matrix).

The AF evolution operator for nonlinear hyperbolic systems is now defined. First, determine a set of predictor characteristic foot locations:

$$x_0^{lk} = x - \frac{\lambda_l + \lambda_k}{2}\Delta t\,, l, k = 1, m \tag{21}$$

Next, determine the predictor values:

$$\boldsymbol{U}^{(l)}(t^{n+1}, x) = \sum_{k=1}^{m} \boldsymbol{R}\boldsymbol{I}_k\boldsymbol{L}\boldsymbol{U}\left(t^n, x_0^{lk}\right)\,, l = 1, m \tag{22}$$

where $\boldsymbol{I_k}$ is a matrix having all elements zero except $\boldsymbol{I_k}(k, k) = 1$. The matrix $\boldsymbol{R}\boldsymbol{I}_k\boldsymbol{L}$ is found in literature as the projector matrix associated with the $k^{th}$ eigenvalue of the system.

Let $\boldsymbol{L}^*$, $\boldsymbol{R}^*$, $\boldsymbol{\Lambda}^*$ be the left and right eigenvectors' matrices and eigenvalues where each eigenvalue and its corresponding eigenvector are evaluated using a different predictor value (e.g., $\lambda_1^* = \lambda_1^*\left(\boldsymbol{U}^{(1)}\right)$, $\lambda_2^* = \lambda_2^*\left(\boldsymbol{U}^{(2)}\right)$, $\boldsymbol{R}_1^* = \boldsymbol{R}_1^*\left(\boldsymbol{U}^{(1)}\right)$, and so on).

The corrector step characteristic foot locations are:

$$x_0^k = x - \lambda_k^* \Delta t\,, k = 1, m \tag{23}$$

Finally, the corrector values are:

$$\boldsymbol{U}(t^{n+1},x) = \sum_{k=1}^{m} \boldsymbol{R}_k^* \, \boldsymbol{L}_k^* \boldsymbol{U}\big(t^n, x_0^k\big) \quad (24)$$

The AF scheme for the Euler equations follows the algorithm presented earlier for the Burgers equation, with some differences. The first difference is that the reconstruction (5) is done independently for each variable. The second difference is that the update of the point values on the cell boundaries $\boldsymbol{U}\big(t^{n+1}, x_{i+1/2}\big)$ and $\boldsymbol{U}\big(t^{n+1/2}, x_{i+1/2}\big)$ is done using (21)-(24).

### 3.3. Sod shock tube problem

The performance of the AF scheme for the Euler equations is investigated using the well-known Sod shock tube problem (Sod, 1978). Consider $x \in [-1\,,1]$ and the following (nondimensional) initial conditions:

$$\begin{cases} \rho = 1\,, p = 1, u = 0 \; for \; x \leq 0 \\ \rho = 0.125\,, p = 0.1\,, u = 0 \; for \; x > 0 \end{cases} \quad (25)$$

The numerical simulation is done using $N = 80$ grid cells, a Courant number $\sigma = 0.7$ for a (nondimensional) maximum time of $t = 0.4$. Figure 3 presents a comparison between the AF scheme results and exact solution of the problem. It can be seen that the AF scheme is capable of capturing the discontinuous solution without difficulties. There are some minor oscillation presents, but it must be remembered that the results shown in Figure 3 were obtained without any limiting or dissipation. The topic of limiting within the context of AF has been explored by Maeng (2017) and Barsukow (2021) but finding a suitable limiter that follows the philosophy of the AF scheme (simple, compact, physically driven) remains an open research question. Nonetheless, good results can be obtained even without any limiting, as shown here.

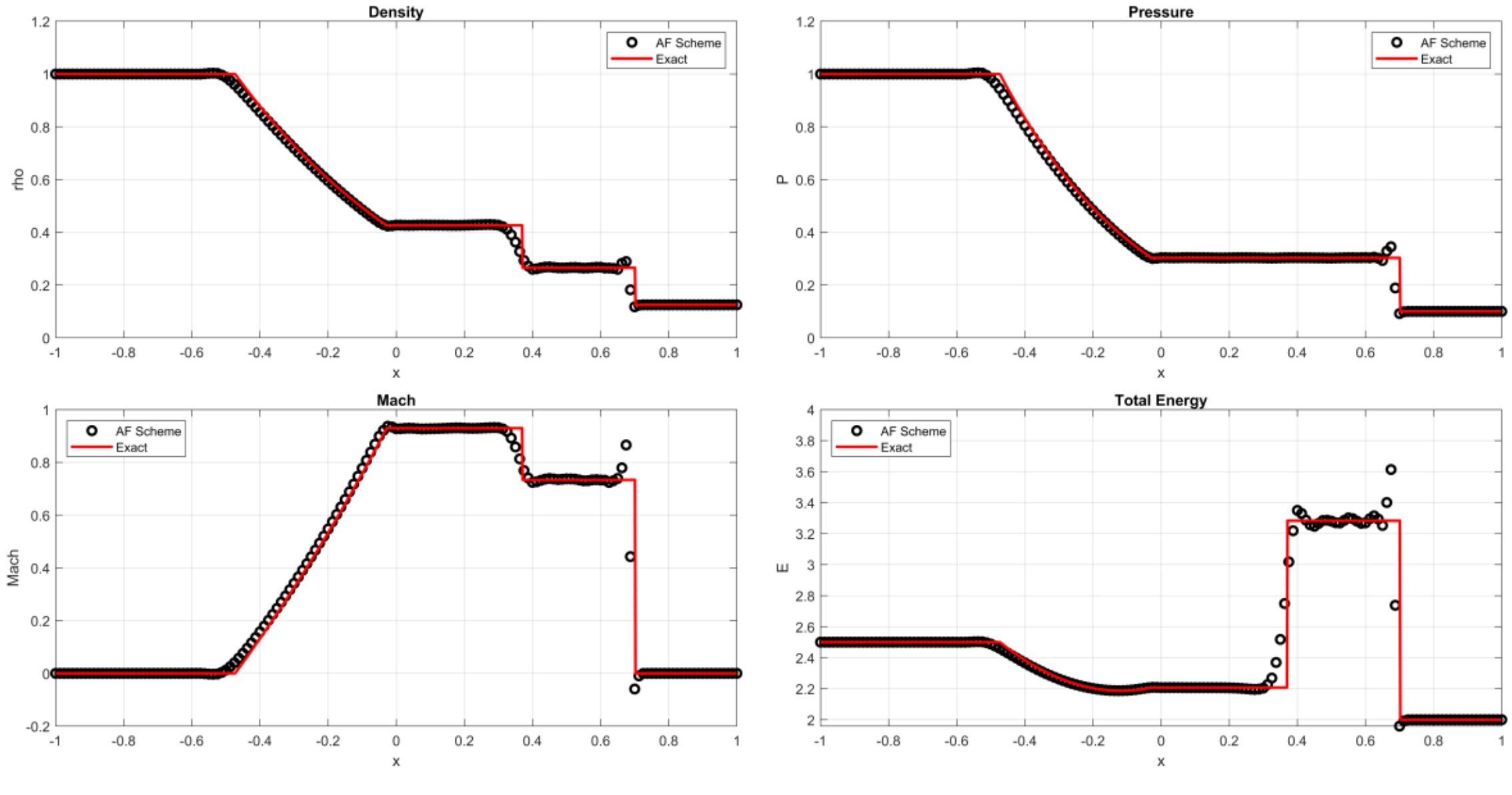

**Figure 3. Sod shock tube problem for the Euler equations. Solution was obtained at $t = 0.4$ with $N = 80$ grid cells and $\sigma = 0.7$. Plotted results include both boundary point values and cell-average values.**

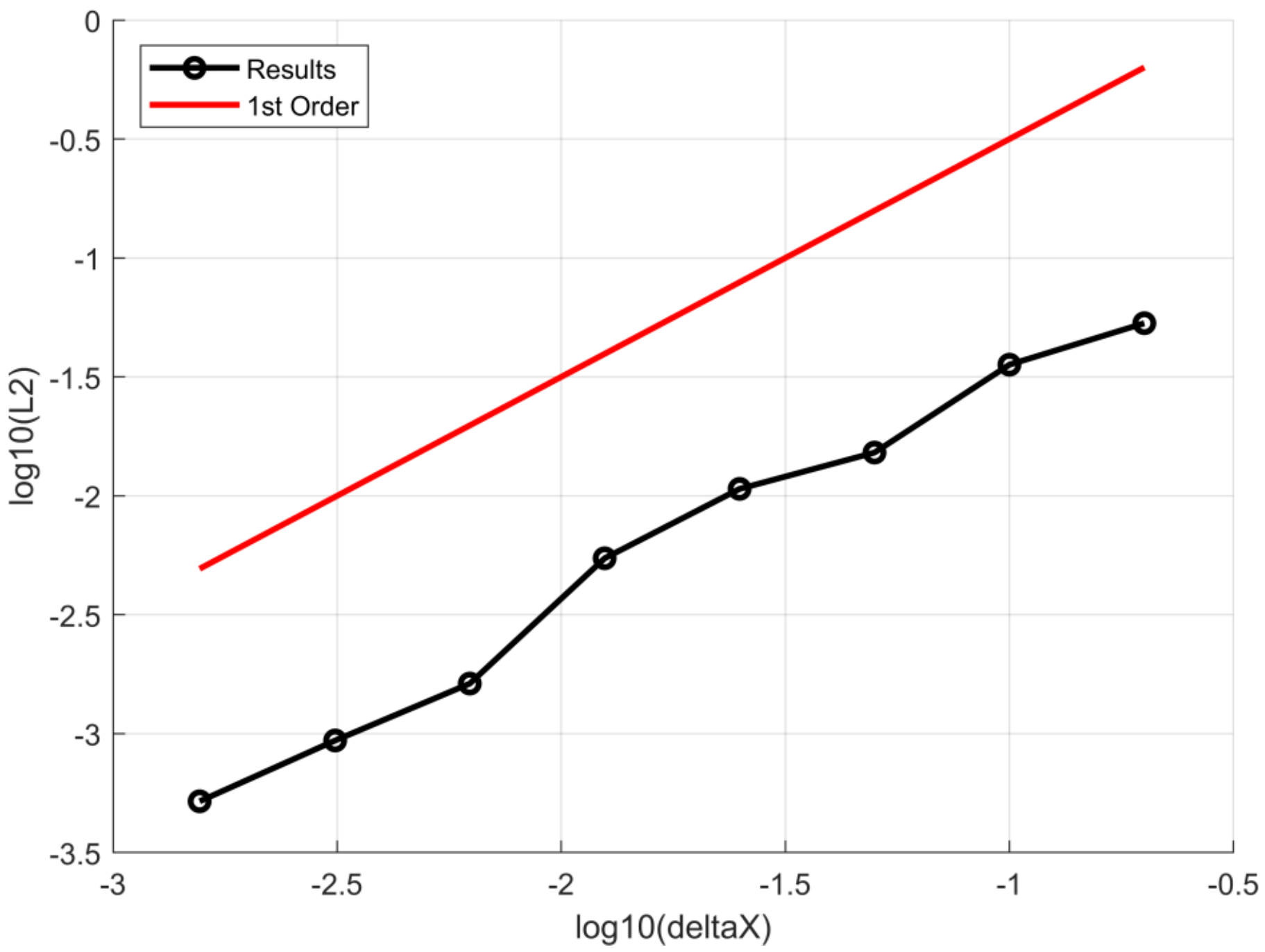

**Figure 4. Order of convergence of the solution to Euler equations at $t = 0.4$ as measured by the $L2$ error with respect to the exact solution. A line corresponding to 1st order is shown for reference.**

Next, the order of convergence is checked. The $L2$ norm of the density error between the solution at increasing grid density and the exact solution is calculated at $t = 0.4$ and plotted in Figure 4. The coarsest grid level has $N = 5$ cells and the number of cells is doubled for each subsequent grid level. As expected, when moving discontinuities are present in the solution, the order of accuracy of the scheme is reduced and the AF scheme converges at first-order accuracy for this test case.

## 4. DIFFUSION AND THE TREATMENT OF SOURCE TERMS

The ability of the AF scheme to handle the diffusion equation is a key step towards application on the Navier-Stokes system. An AF scheme has been developed and applied to the diffusion equation in the work of Nishikawa et al. (2014) and Nishikawa and Roe (2016). The scheme used in the present work is different compared to the one employed by the mentioned authors, but the first-order hyperbolic system formulation of the diffusion equation is kept. This formulation was developed by Nishikawa (2007) and is equivalent to the diffusion equation in the limit of vanishing relaxation time.

### 4.1. Hyperbolic formulation of the diffusion equation

Consider the one-dimensional diffusion equation:

$$\frac{\partial U}{\partial t} = \frac{\partial}{\partial x}\left(\nu \frac{\partial U}{\partial x}\right) \tag{26}$$

where $\nu$ is a (typically small) diffusion coefficient.

Pass to the hyperbolic system formulation of Nishikawa (2007):

$$\begin{cases} \dfrac{\partial U}{\partial t} - \dfrac{\partial P}{\partial x} = 0 \\ \dfrac{\partial P}{\partial t} - \dfrac{\nu}{T}\dfrac{\partial U}{\partial x} = -\dfrac{\nu}{T}\dfrac{P}{\nu} \end{cases} \tag{27}$$

Here, $P = \nu \partial U/\partial x$ and $T$ is the relaxation time. The system (27) is equivalent to (26) in the limit of vanishing relaxation time. For steady-state solutions, $T$ can be chosen larger, but for time-accurate unsteady solution, $T$ must be carefully chosen and sufficiently small. For best results, dual time stepping methods are recommended (Mazaheri and Nishikawa, 2014). The advantage of (27) is the hyperbolic nature of the system, meaning that AF schemes developed for advection-dominated phenomena can be extended to diffusion as well. The disadvantage will be dealing with the stiff source term caused by the small relaxation time $T$.

The system (27) can be written as a typical balance law (conservation law with source term):

$$\frac{\partial \boldsymbol{U}}{\partial t} + \boldsymbol{P}\frac{\partial \boldsymbol{F}}{\partial x} = \boldsymbol{PS} \tag{28}$$

with $\boldsymbol{P}$ acting like a pre-conditioning matrix.

The various terms are:

$$\boldsymbol{U} = \begin{bmatrix} U \\ P \end{bmatrix} \quad \boldsymbol{F} = \begin{bmatrix} -P \\ -U \end{bmatrix} \quad \boldsymbol{S} = \begin{bmatrix} 0 \\ -\dfrac{P}{\nu} \end{bmatrix} \quad \boldsymbol{P} = \begin{bmatrix} 1 & 0 \\ 0 & \dfrac{\nu}{T} \end{bmatrix} \tag{29}$$

For ease of notation, the following variable is introduced:

$$a_\nu = \sqrt{\frac{\nu}{T}} \tag{30}$$

Also, to allow for verification solutions to be easily obtained, an additional source term $S_1$ is added, its exact shape being defined in the application stage:

$$\boldsymbol{PS} = \begin{bmatrix} S_1 \\ -\dfrac{P}{T} \end{bmatrix} \tag{31}$$

The eigenvalues and eigenvectors of this system can be determined without any difficulties:

$$\boldsymbol{\Lambda} = \begin{bmatrix} a_\nu & 0 \\ 0 & -a_\nu \end{bmatrix} \quad \mathbf{R} = \begin{bmatrix} 1 & 1 \\ -a_\nu & a_\nu \end{bmatrix} \quad \mathbf{L} = \begin{bmatrix} \dfrac{1}{2} & -\dfrac{1}{2a_\nu} \\ \dfrac{1}{2} & \dfrac{1}{2a_\nu} \end{bmatrix} \tag{32}$$

The information in (32) is sufficient to apply the AF predictor-corrector evolution operator defined in (21)-(24). However, the source term must also be taken into consideration before any solution is attempted.

## 4.2. Evolution operator with source term

First, the source term must be included in the evolution operator for the boundary point values. By multiplying (28) to the left by $\boldsymbol{L}$ and defining the characteristic variables as $\mathbf{W} = \boldsymbol{LU}$, the characteristic system is obtained:

$$\frac{\partial \mathbf{W}}{\partial t} + \boldsymbol{\Lambda}\frac{\partial \mathbf{W}}{\partial x} = \boldsymbol{S^w} \tag{33}$$

where $\boldsymbol{S^w} = \boldsymbol{LPS}$.

The diffusion system being linear, the characteristic variables can be explicitly determined:

$$\mathbf{W} = \begin{bmatrix} W_1 \\ W_2 \end{bmatrix} = \begin{bmatrix} \dfrac{U}{2} - \dfrac{P}{2a_v} \\ \dfrac{U}{2} + \dfrac{P}{2a_v} \end{bmatrix} \tag{34}$$

The values of the characteristic variables $W_k$ can be obtained by integrating the ordinary differential equation $dW_k/dt = S_k^w$ along the characteristic curve $C^k = x - \lambda_k \Delta t$ for one time step to get:

$$\int_{W_k(t^n, x-\lambda_k\Delta t)}^{W_k(t^{n+1}, x)} \frac{dW_k}{S_k^w(W_1, W_2)} = \Delta t\,, k = 1,2 \tag{35}$$

It is seen that $S_k^w$ is a linear function of the characteristic variables $W_k$:

$$\boldsymbol{S^w} = \begin{bmatrix} \dfrac{S_1}{2} - \dfrac{P}{2a_v T} \\ \dfrac{S_1}{2} + \dfrac{P}{2a_v T} \end{bmatrix} = \begin{bmatrix} \dfrac{S_1}{2} + \dfrac{1}{2T}(W_1 - W_2) \\ \dfrac{S_1}{2} + \dfrac{1}{2T}(W_2 - W_1) \end{bmatrix} \tag{36}$$

It is possible to analytically determine the integral in (35) and obtain an exact evolution equation for the characteristic variables:

$$\begin{aligned} W_1(t^{n+1}, x) &= W_1(t^n, x - \lambda_k \Delta t) e^{\frac{\Delta t}{2T}} + \left(1 - e^{\frac{\Delta t}{2T}}\right)(W_2(t^n, x) - S_1 T) \\ W_2(t^{n+1}, x) &= W_2(t^n, x - \lambda_k \Delta t) e^{\frac{\Delta t}{2T}} + \left(1 - e^{\frac{\Delta t}{2T}}\right)(W_1(t^n, x) - S_1 T) \end{aligned} \tag{37}$$

The above approach in which each characteristic equation is integrated separately might not work well for nonlinear system of equations where the characteristic variables can only be defined in the more general sense $\partial\mathbf{W}$ and where eigenvalues are a function of all characteristic variables.

As the starting point for a more general alternative, the characteristic variables at a point are given by the values of the characteristic variables at the foot of the characteristic curve $C^k$ plus the effect of the source term $S_k^w$ integrated for one time step between the foot of the characteristic curve and the point of interest. This starting point is still done separately for each characteristic equation but following steps will rely on a modification of the predictor-corrector scheme that was shown to work well for equations without source terms. We have:

$$\partial W_k(t^{n+1},x) = \partial W_k(t^n, x-\lambda_k\Delta t) + \int_0^{\Delta t} S_k^w dt \ , k = 1,2 \tag{38}$$

The integral over time from the source term can be transformed into an integral over space via a change of the integration variable knowing that along a characteristic curve $x - \lambda_k t = constant$:

$$\partial W_k(t^{n+1},x) = \partial W_k(t^n, x-\lambda_k\Delta t) + \int_{x-\lambda_k\Delta t}^{x} \frac{S_k^w(t^n,x)}{\lambda_k} dx \ , k = 1,2 \tag{39}$$

The integral can be approximated by any suitable quadrature formula, for example Simpson's 1/3 quadrature for 3rd order accuracy. For nonlinear systems of equations, the eigenvalues appearing under the integral must be calculated at the appropriate instance in time and coordinate:

$$\begin{gathered}\int_{x-\lambda_k\Delta t}^{x} S_k^w(t^n,x)dx \\ \cong \frac{\lambda_k(t^n,x-\lambda_k\Delta t)\Delta t}{6}\left(\frac{S_k^w(t^n,x-\lambda_k\Delta t)}{\lambda_k(t^n,x-\lambda_k\Delta t)} + 4\frac{S_k^w\left(t^n,x-\lambda_k\frac{\Delta t}{2}\right)}{\lambda_k\left(t^n,x-\lambda_k\frac{\Delta t}{2}\right)} + \frac{S_k^w(t^n,x)}{\lambda_k(t^n,x)}\right)\end{gathered} \tag{40}$$

Where $\lambda_k(t^n, x-\lambda_k\Delta t)$ is a notation for $\lambda_k(W_1(t^n,x-\lambda_k\Delta t), W_2(t^n,x-\lambda_k\Delta t),\dots)$ and similarly for the other terms.

With the starting point available, it follows that:

$$\begin{aligned}\partial \boldsymbol{U}(t^{n+1},x) &= \boldsymbol{L}^{-1}\partial\mathbf{W}(t^{n+1},x) = \boldsymbol{R}\partial\mathbf{W}(t^{n+1},x) \\ &= \sum_{k=1}^{m} \boldsymbol{R}_k\,\partial W_k(t^{n+1},x) \\ &= \sum_{k=1}^{m} \boldsymbol{R}_k\left[\partial W_k(t^n,x-\lambda_k\Delta t) + \int_{x-\lambda_k\Delta t}^{x}\frac{S_k^w}{\lambda_k}dx\right] \\ &= \sum_{k=1}^{m} \boldsymbol{R}_k\Bigg[\partial W_k(t^n,x-\lambda_k\Delta t) \\ &\quad + \frac{\lambda_k\Delta t}{6}\left(\frac{S_k^w(t^n,x-\lambda_k\Delta t)}{\lambda_k(t^n,x-\lambda_k\Delta t)} + 4\frac{S_k^w\left(t^n,x-\lambda_k\frac{\Delta t}{2}\right)}{\lambda_k\left(t^n,x-\lambda_k\frac{\Delta t}{2}\right)} + \frac{S_k^w(t^n,x)}{\lambda_k(t^n,x)}\right)\Bigg] \\ &= \sum_{k=1}^{m}\Bigg[\boldsymbol{R}_k\boldsymbol{L}_k\partial\boldsymbol{U}(t^n,x-\lambda_k\Delta t) + \frac{\lambda_k\Delta t}{6}\boldsymbol{R}_k\boldsymbol{L}_k\boldsymbol{Q}(t^n,x-\lambda_k\Delta t) \\ &\quad + \frac{2\lambda_k\Delta t}{3}\boldsymbol{R}_k\boldsymbol{L}_k\boldsymbol{Q}\left(t^n,x-\lambda_k\frac{\Delta t}{2}\right) + \frac{\lambda_k\Delta t}{6}\boldsymbol{R}_k\boldsymbol{L}_k\boldsymbol{Q}(t^n,x)\Bigg]\end{aligned} \tag{41}$$

In the above, the notation $Q_k = S_k/\lambda_k$ was introduced.

The AF predictor-corrector evolution operator defined in (21)-(24) can now be updated to take into consideration source terms. First, determine a set of predictor characteristic foot locations:

$$x_0^{lk} = x - \frac{\lambda_l + \lambda_k}{2}\Delta t\,, l, k = 1, m$$
$$x2_0^{lk} = x - \frac{\lambda_l + \lambda_k}{2}\frac{\Delta t}{2}\,, l, k = 1, m \tag{42}$$

Next, determine the predictor values:

$$\boldsymbol{U}^{(l)}(t^{n+1}, x) = \sum_{k=1}^{m}\Big[\boldsymbol{RI}_k\boldsymbol{LU}(t^n, x_0^{lk}) + \frac{\lambda_k\Delta t}{6}\boldsymbol{RI}_k\boldsymbol{LQ}(t^n, x_0^{lk})$$
$$+ \frac{2\lambda_k\Delta t}{3}\boldsymbol{RI}_k\boldsymbol{LQ}(t^n, x2_0^{lk}) + \frac{\lambda_k\Delta t}{6}\boldsymbol{RI}_k\boldsymbol{LQ}(t^n, x)\Big]\,, l = 1, m \tag{43}$$

The corrector step characteristic foot locations are:

$$x_0^k = x - \lambda_k^*\Delta t\,, k = 1, m$$
$$x2_0^k = x - \lambda_k^*\frac{\Delta t}{2}\,, k = 1, m \tag{44}$$

Finally, the corrector values are:

$$\boldsymbol{U}(t^{n+1}, x) = \Bigg[\sum_{k=1}^{m}\boldsymbol{R}_k^*\,\boldsymbol{L}_k^*\boldsymbol{U}(t^n, x_0^k) + \frac{\lambda_k^*\Delta t}{6}\boldsymbol{R}_k^*\boldsymbol{L}_k^*\boldsymbol{Q}(t^n, x_0^k)$$
$$+ \frac{2\lambda_k^*\Delta t}{3}\boldsymbol{R}_k^*\boldsymbol{L}_k^*\boldsymbol{Q}(t^n, x2_0^k) + \frac{\lambda_k^*\Delta t}{6}\boldsymbol{R}_k^*\boldsymbol{L}_k^*\boldsymbol{Q}(t^n, x)\Bigg] \tag{45}$$

### 4.3. Inclusion of source term in conservative update

The second aspect that must be updated to account for the source term is the cell-average balance law. This is written as:

$$\overline{\boldsymbol{U}}_i^{n+1} - \overline{\boldsymbol{U}}_i^n = -\frac{\Delta t}{\Delta x}\left(\overline{\boldsymbol{F}}(x_{i+1/2}) - \overline{\boldsymbol{F}}(x_{i-1/2})\right) + \frac{1}{\Delta x}\int_{t^n}^{t^{n+1}}\int_{x_{i-1/2}}^{x_{i+1/2}}\boldsymbol{S}dx\,dt \tag{46}$$

Using Simpson's 1/3 quadrature rule to approximate the integrals is not a viable option since there are not enough quadrature points available. More specifically, only the cell-average value at the current time $\overline{\boldsymbol{U}}_i^n$ is known, but not at $\overline{\boldsymbol{U}}_i^{n+1/2}$ or $\overline{\boldsymbol{U}}_i^{n+1}$ as required by Simpson's rule.

However, a reconstruction can be defined to make use of all cell data available, including the updated values on the cell boundaries $\boldsymbol{U}(t^{n+1}, x_{i+1/2})$, $\boldsymbol{U}(t^{n+1/2}, x_{i+1/2})$ and so on. This will make the reconstruction used for the source terms different compared to the reconstruction (5) used for other aspects of the scheme.

It must also be mentioned that the source term in (46) is linear:

$$\int_{t^n}^{t^{n+1}}\int_{x_{i-1/2}}^{x_{i+1/2}}\boldsymbol{S}dx\,dt = \boldsymbol{K}\int_{t^n}^{t^{n+1}}\int_{x_{i-1/2}}^{x_{i+1/2}}\boldsymbol{U}dx\,dt \tag{47}$$

with $\boldsymbol{K}$ being a constant term.

The reconstruction was obtained by Barsukow et al. (2021) and it is obtained as:

$$\int_{t^n}^{t^{n+1}} \int_{x_{i-1/2}}^{x_{i+1/2}} U dx\, dt$$
$$= \Delta x \Delta t \left( \bar{U}_i^n + \frac{1}{12} U_{i-1/2}^{n+1} + \frac{1}{3} U_{i-1/2}^{n+1/2} - \frac{5}{12} U_{i-1/2}^{n} + \frac{1}{12} U_{i+1/2}^{n+1} + \frac{1}{3} U_{i+1/2}^{n+1/2} - \frac{5}{12} U_{i+1/2}^{n} \right) \quad (48)$$

By using (48) and (47), the source term value in every cell can be determined before being used in the conservative update (46).

### 4.4. Time accurate calculations via dual time stepping

As mentioned, the hyperbolic system (27) is equivalent to the diffusion equation (26) in the limit of vanishing relaxation time. This can be achieved, at each physical time step, by converging (27) to a steady-state solution in a dual time formulation (Jameson, 1991).

Consider that the balance law (28) is discretised using the following backward difference formula for the time derivative:

$$\frac{\alpha}{\Delta t} \bar{\boldsymbol{U}}_i^{n+1} + \frac{\beta}{\Delta t} \bar{\boldsymbol{U}}_i^{n} + \frac{\gamma}{\Delta t} \bar{\boldsymbol{U}}_i^{n-1} + \frac{\delta}{\Delta t} \bar{\boldsymbol{U}}_i^{n-2} = -\frac{1}{\Delta x}\left( \bar{\boldsymbol{F}}(x_{i+1/2}) - \bar{\boldsymbol{F}}(x_{i-1/2}) \right) + \frac{1}{\Delta t \Delta x} \int_{t^n}^{t^{n+1}} \int_{x_{i-1/2}}^{x_{i+1/2}} \boldsymbol{S} dx\, dt \quad (49)$$

A first-order accurate in time formula has:

$$\alpha = 1, \quad \beta = -1, \quad \gamma = 0, \quad \delta = 0 \quad (50)$$

A second-order accurate in time formula has:

$$\alpha = 3/2, \quad \beta = -2, \quad \gamma = 1/2, \quad \delta = 0 \quad (51)$$

A third-order accurate in time formula has:

$$\alpha = 11/6, \quad \beta = -3, \quad \gamma = 3/2, \quad \delta = -1/3 \quad (52)$$

Following Jameson (1991), define the modified residual as:

$$\boldsymbol{R}^* = -\frac{1}{\Delta x}\left( \bar{\boldsymbol{F}}(x_{i+1/2}) - \bar{\boldsymbol{F}}(x_{i-1/2}) \right) + \frac{1}{\Delta t \Delta x} \int_{t^n}^{t^{n+1}} \int_{x_{i-1/2}}^{x_{i+1/2}} \boldsymbol{S} dx\, dt - \frac{\beta}{\Delta t} \bar{\boldsymbol{U}}_i^{n} - \frac{\gamma}{\Delta t} \bar{\boldsymbol{U}}_i^{n-1} - \frac{\delta}{\Delta t} \bar{\boldsymbol{U}}_i^{n-2} \quad (53)$$

The dual time problem is formulated as:

$$\frac{d\bar{\boldsymbol{U}}_i}{d\tau} = -\frac{\alpha}{\Delta t} \bar{\boldsymbol{U}}_i + \boldsymbol{R}^* \quad (54)$$

This can now be time-marched towards steady state with a dual time step $\Delta\tau$. The terms due to discretisation of the physical time derivative are kept constant during the dual time marching towards steady state. Additionally, if an explicit dual time scheme is used, the first term on the right-hand side of (54) must be discretised implicitly to ensure adequate stability:

$$\overline{\boldsymbol{U}}_i^{k+1} - \overline{\boldsymbol{U}}_i^{k} = -\frac{\alpha\Delta\tau}{\Delta t}\overline{\boldsymbol{U}}_i^{k+1} + \Delta\tau\boldsymbol{R}^* \quad (55)$$

Once the dual time problem (55) has been converged to steady state, the solution at the next physical time step is obtained:

$$\overline{\boldsymbol{U}}_i^{n+1} = \overline{\boldsymbol{U}}_i^{k+1} \quad (56)$$

### 4.5. Boundary conditions

In the AF method, just like in many FV solvers, it may be the most convenient approach to let the solver determine the numerical flux through the domain boundaries using (11). Assuming $N = 1$ is the first cell adjacent to the left side of the domain, Dirichlet boundary condition for the primitive or conservative variables can be specified in a fictitious ghost cell $N = 0$, and the flux through the boundary is given by:

$$F_{1/2} = \frac{\Delta t}{6}\left(F\left(U(t^n, x_{1/2})\right) + 4F\left(U(t^{n+1/2}, x_{1/2})\right) + F\left(U(t^{n+1}, x_{1/2})\right)\right) \quad (57)$$

where $x_1$ is the coordinate of the first cell centre, and $x_{1/2}$ and $x_{3/2}$ are the coordinates of the left and right faces of the first cell.

To use (57), the conservative variables must be evolved on the boundary face. However, when source terms are present in the equation, their contribution must be added only from points that are inside the physical domain.

The eigenvalues of the diffusion equation are:

$$\begin{bmatrix} a_v \\ -a_v \end{bmatrix} \quad (58)$$

The first eigenvalue is positive, and its associated characteristic curve will enter the domain at the left boundary $x_{1/2}$. The second eigenvalue is negative, and its associated characteristic curve will leave the domain at the same boundary.

The first characteristic variable is kept constant along the corresponding characteristic curve. Since this characteristic curve brings information from outside the domain (from the ghost cell), the effect of the source term should not be included, giving:

$$W_1(t^{n+1}, x_{1/2}) = W_1(t^n, x_{1/2} - \lambda_1\Delta t) \quad (59)$$

Or, writing explicitly,

$$\frac{U_{1/2}^{n+1}}{2} - \frac{P_{1/2}^{n+1}}{2a_v} = \frac{U_0}{2} - \frac{P_0}{2a_v} \quad (60)$$

Here, by $U_0$ and $P_0$, we denote the conservative variable values inside the ghost cell, assumed to be constant in time and provided via Dirichlet boundary conditions.

The second characteristic curve brings information from inside the domain, so the effect of the source term must be included:

$$W_2(t^{n+1}, x_{1/2}) = W_2(t^n, x_{1/2} - \lambda_2\Delta t) + \int_{x_{1/2}-\lambda_2\Delta t}^{x_{1/2}} \frac{S_2^w(t^n, x)}{\lambda_2} dx \quad (61)$$

The integral can be determined as explained previously, using data from the first cell of the domain. For ease of notation, let us denote:

$$\int_{x_{1/2}-\lambda_k\Delta t}^{x_{1/2}} \frac{S_2^w(t^n,x)}{\lambda_2} dx = I_2 \tag{62}$$

Then,

$$\frac{U_{1/2}^{n+1}}{2} + \frac{P_{1/2}^{n+1}}{2a_v} = \frac{U_d}{2} - \frac{P_d}{2a_v} + I_2 \tag{63}$$

Here, $U_d$ and $P_d$ are $U$ and $P$ calculated via reconstruction at the foot of the characteristic curve, $d = x_{1/2} - \lambda_2\Delta t$, with data from time level $t^n$.

The equations (60) and (63) can be solved for the conservative variables' values on the boundary $x_{1/2}$:

$$\begin{gathered} P_{1/2}^{n+1} = \frac{1}{2}(P_d + P_0) + \frac{a_v}{2}(U_d - U_0) + a_v I_2 \\ U_{1/2}^{n+1} = U_0 + \frac{1}{a_v}\left(P_{1/2}^{n+1} - P_0\right) \end{gathered} \tag{64}$$

Determination of the conservative variables on the boundary then allows for the flux through the boundary to be calculated using (57).

A similar approach is used for right boundary of the domain, $x_{N+1/2}$. Knowing that at the right boundary the first characteristic curve leaves the domain, and the second characteristic curve enters the domain from the ghost cell $N+1$ where the Dirichlet boundary conditions are provided, the following equations are obtained for the conservative variables' values on this boundary:

$$\begin{gathered} P_{N+1/2}^{n+1} = \frac{1}{2}(P_d + P_0) - \frac{a_v}{2}(U_d - U_0) - a_v I_1 \\ U_{N+1/2}^{n+1} = U_0 - \frac{1}{a_v}\left(P_{N+1/2}^{n+1} - P_0\right) \end{gathered} \tag{65}$$

### 4.6. Application to the diffusion equation

With all elements of the AF scheme defined, an unsteady example can be considered. For the diffusion equation (28), consider the additional source term:

$$S_1 = 0.1e^{0.05x} \tag{66}$$

The equation is solved in $x \in [-3\pi/2\,, 3\pi/2]$ with $\nu = 0.01$. The Dirichlet boundary conditions are:

$$\begin{gathered} U(t, -3\pi/2) = 0.1e^{-0.05\frac{3\pi}{2}} \\ U(t, 3\pi/2) = 0.1e^{0.05\frac{3\pi}{2}} \end{gathered} \tag{67}$$

And the initial condition is:

$$U(0,x) = \cos x \tag{68}$$

Following Nishikawa (2007), the relaxation time is set as $T = L^2/\nu$, $L = 1/2\pi$. The numerical simulation is done using $N = 25$ grid cells, a physical time step $\Delta t = 0.1\ s$, a dual time Courant number $\sigma = 0.9$ and dual time steady state convergence criteria of $10^{-10}$. The simulation is run up to a total physical time of $20\ s$.

Figure 5 presents a comparison between the AF scheme results and the results obtained for the same problem using a 4th order accurate in space and 3rd order accurate in time finite-difference (FD) solution, run on $N = 2560$ grid points. The finite-difference solution is taken as a substitute "exact" solution, and the $L2$ norm of the error between the AF solution at increasing grid density and the FD solution is calculated at $t = 20\ s$ and plotted in Figure 6. The coarsest grid level has $N = 5$ cells and the number of cells is doubled for each subsequent grid level. It can be observed that the AF scheme achieves convergence at third-order accuracy for the diffusion equation.

## 5. PROGRESS TOWARDS SOLVING THE NAVIER-STOKES EQUATIONS

The one-dimensional Navier-Stokes equations have the following shape:

$$\frac{\partial \boldsymbol{U}}{\partial t} + \frac{\partial \boldsymbol{F}}{\partial x} = \boldsymbol{0} \tag{69}$$

The various terms are:

$$\boldsymbol{U} = \begin{bmatrix} \rho \\ \rho u \\ \rho E \end{bmatrix} \quad \boldsymbol{F} = \begin{bmatrix} \rho u \\ \rho u^2 + p - \tau \\ \rho u H - \tau u + q \end{bmatrix} \tag{70}$$

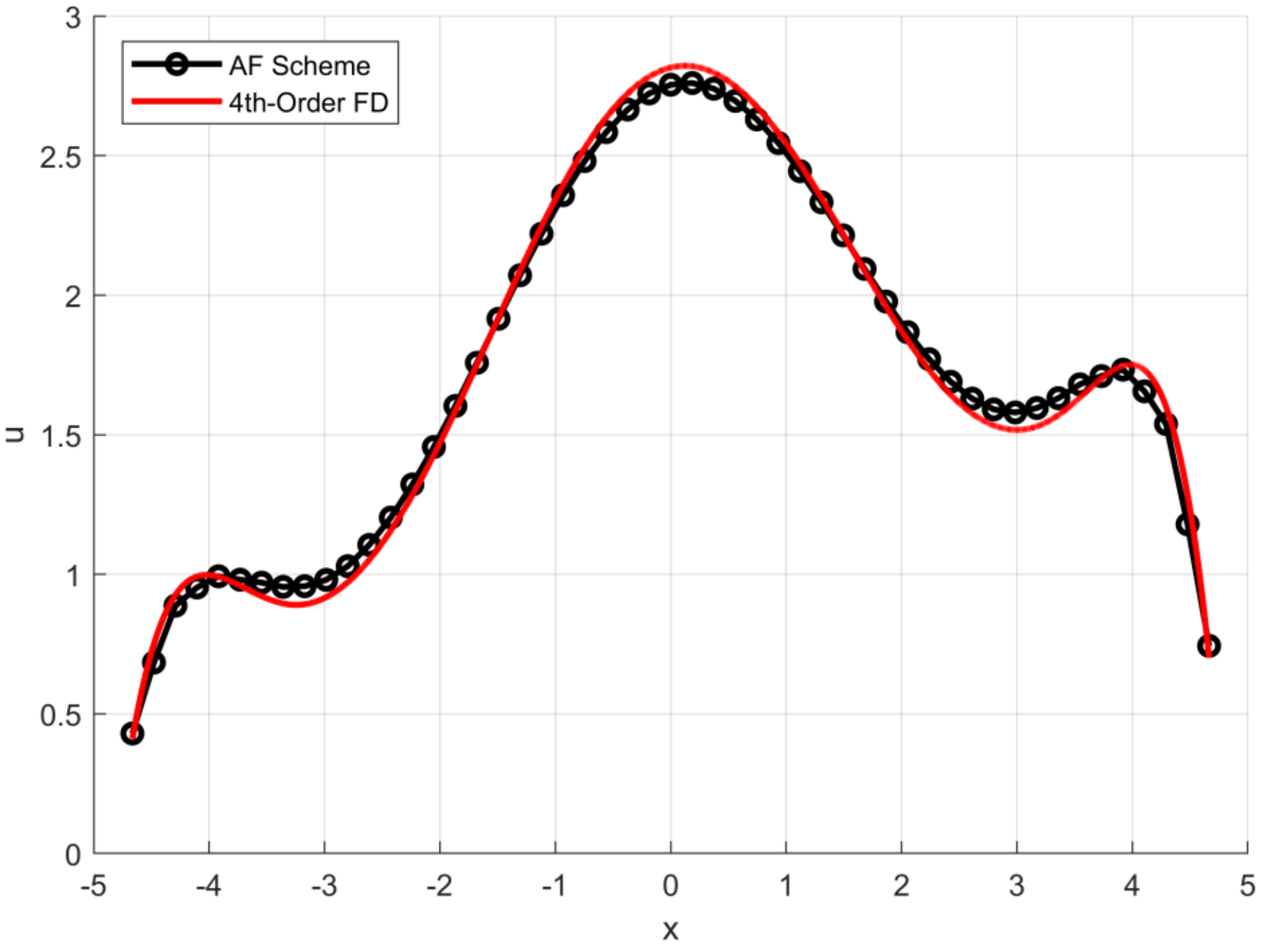

**Figure 5. Unsteady solution using hyperbolic system form of the diffusion equation obtained using AF and 4th order FD. Plotted results for AF include both boundary point values and cell-average values.**

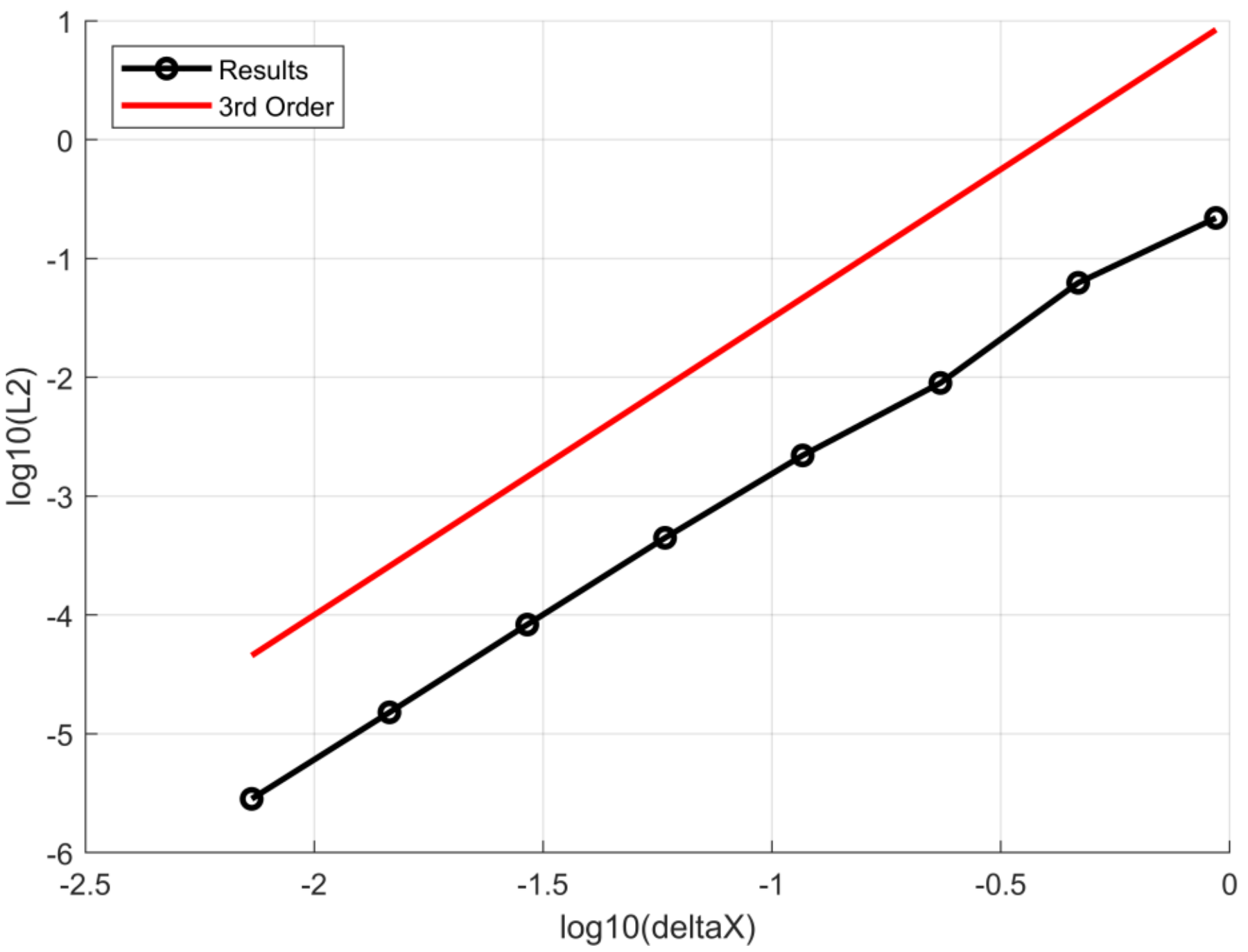

**Figure 6. Order of convergence of the unsteady solution of the hyperbolic system form of the diffusion equation measured by the $L2$ error with respect to the FD solution. A line corresponding to 3rd order is shown for reference.**

With:

$$\tau = \frac{4}{3}\mu\frac{\partial u}{\partial x}$$
$$q = -\frac{\gamma\mu}{Pr(\gamma-1)}\frac{\partial}{\partial x}\left(\frac{p}{\rho}\right) \tag{71}$$

Here, $\rho$ is the density, $u$ is the flow velocity, $E$ is the specific total energy, $p$ is the pressure and $H = E + p/\rho$ is specific total enthalpy, $\tau$ is the viscous stress and $q$ is the heat flux. It is assumed the fluid also satisfies the equation of state for ideal gases, which can be written in a convenient form as $p = (\gamma - 1)(\rho E - \rho u^2/2)$, with $\gamma$ being the ratio of specific heats, having a value $\gamma = 1.4$ for air. The dynamic viscosity $\mu$ is assumed to be a function of temperature only, typically expressed via Sutherland's law, while the Prandtl number for air is $Pr = 0.72$.

### 5.1. Hyperbolic formulation of the Navier-Stokes equations

Similar to the approach used for the diffusion equation, the Navier-Stokes equations are reformulated as a first-order hyperbolic system. The hyperbolic formulation is that of Nishikawa (2011).

Let:

$$\frac{\partial\tau}{\partial t} = \frac{\mu_v}{T_v}\frac{\partial u}{\partial x} - \frac{\tau}{T_v}$$
$$\frac{\partial q}{\partial t} = -\frac{\mu_h}{T_h}\frac{1}{\gamma-1}\frac{\partial}{\partial x}\left(\frac{p}{\rho}\right) - \frac{q}{T_h} \tag{72}$$

where $\mu_v = 4\mu/3$, $\mu_h = \gamma\mu/Pr$, while $T_v$ and $T_h$ are relaxation times. As previously discussed for the diffusion equation, the hyperbolic reformulation is equivalent to (70) in the limit of vanishing relaxation time.

The hyperbolic formulation of the Navier-Stokes equations has the following shape:

$$\frac{\partial \boldsymbol{U}}{\partial t} + \boldsymbol{P}\frac{\partial \boldsymbol{F}}{\partial x} = \boldsymbol{PS} \tag{73}$$

with $\boldsymbol{P}$ acting like a pre-conditioning matrix.

The various terms are:

$$\boldsymbol{U} = \begin{bmatrix} \rho \\ \rho u \\ \rho E \\ \tau \\ q \end{bmatrix} \quad \boldsymbol{F} = \begin{bmatrix} \rho u \\ \rho u u + p - \tau \\ \rho u H - \tau u + q \\ -u \\ \frac{1}{\gamma - 1}\frac{p}{\rho} \end{bmatrix} \quad \boldsymbol{S} = \begin{bmatrix} 0 \\ 0 \\ 0 \\ -\frac{\tau}{\mu_V} \\ -\frac{q}{\mu_H} \end{bmatrix} \quad \boldsymbol{P} = \begin{bmatrix} 1 & 0 & 0 & 0 & 0 \\ 0 & 1 & 0 & 0 & 0 \\ 0 & 0 & 1 & 0 & 0 \\ 0 & 0 & 0 & \frac{\mu_v}{T_v} & 0 \\ 0 & 0 & 0 & 0 & \frac{\mu_h}{T_h} \end{bmatrix} \tag{74}$$

The main challenge related to applying the evolution operator presented previously is the calculation of the eigenvalues and eigenvectors of the Jacobian matrix $\boldsymbol{A} = \partial \boldsymbol{F}/\partial \boldsymbol{U}$. The full Jacobian has not been analysed in Nishikawa (2011), nor in subsequent works on the topic (for example, Li et al., 2018). Indeed, determining an analytical solution for the eigenvalues of the full Jacobian of the system still proves elusive. Some notes are nonetheless presented here for the completeness of the presentation, and in hopes of further work being carried out.

### 5.2. Jacobian matrix eigenvalues

The Jacobian matrix in conservative variables is given by:

$$\boldsymbol{A} = \boldsymbol{P}\frac{\partial \boldsymbol{F}}{\partial \boldsymbol{U}} = \begin{bmatrix} 0 & 1 & 0 & 0 & 0 \\ \frac{\gamma - 3}{2}u^2 & (3-\gamma)u & \gamma - 1 & -1 & 0 \\ -\gamma u E + (\gamma - 1)u^3 + \frac{u\tau}{\rho} & \gamma E - \frac{3}{2}(\gamma - 1)u^2 - \frac{\tau}{\rho} & \gamma u & -u & 1 \\ a_v^2 u & -a_v^2 & 0 & 0 & 0 \\ a_h^2(u^2 - E) & -a_h^2 u & a_h^2 & 0 & 0 \end{bmatrix} \tag{75}$$

In primitive variables it is:

$$\boldsymbol{B} = \boldsymbol{P}\frac{\partial \boldsymbol{F}}{\partial \boldsymbol{Q}} = \begin{bmatrix} u & \rho & 0 & 0 & 0 \\ 0 & u & \frac{1}{\rho} & -\frac{1}{\rho} & 0 \\ 0 & \rho a^2 - (\gamma - 1)\tau & u & 0 & \gamma - 1 \\ 0 & -\rho a_v^2 & 0 & 0 & 0 \\ -\frac{a_h^2 a^2}{\gamma(\gamma - 1)} & 0 & \frac{a_h^2}{\gamma - 1} & 0 & 0 \end{bmatrix} \tag{76}$$

With $a$ being the speed of sound, $\boldsymbol{Q} = [\rho, u, p, \tau, q]^T$ being the primitive variables and the following notations being introduced:

$$a_v = \sqrt{\frac{\mu_v/\rho}{T_v}} = \sqrt{\frac{\nu_v}{T_v}} \quad a_h = \sqrt{\frac{\mu_h/\rho}{T_h}} = \sqrt{\frac{\nu_h}{T_h}} \tag{77}$$

It is possible to obtain the characteristic polynomial of the primitive Jacobian:

$$\begin{gathered} |\boldsymbol{B} - \lambda \boldsymbol{I}| = 0 \\ (u-\lambda)c_1 + \lambda(u-\lambda)^2 c_2 + \lambda c_3 + \lambda^2(u-\lambda)^3 - \lambda^2(u-\lambda)c_4 = 0 \end{gathered} \tag{78}$$

The various terms are:

$$\begin{gathered} c_1 = a_v^2 a_h^2 \\ c_2 = a_h^2 + a_v^2 \\ c_3 = -\frac{a_h^2 a^2}{\gamma} \\ c_4 = a^2 - (\gamma - 1)\frac{\tau}{\rho} \end{gathered} \tag{79}$$

So far, any attempts at analytically determining the roots of (78) have proven unsuccessful. Assuming a dual time approach is used, where the relaxation times can be defined similar to the work done on the diffusion equation, $T_v = L^2/\nu_v$ and $T_h = L^2/\nu_h$, $L = 1/2\pi$, an order of magnitude investigation of (76) shows that the terms $-\rho a_v^2$ and $a_h^2/(\gamma - 1)$ are both $\mathcal{O}(10^{-10})$ and at least 4 order of magnitude lower compared to all other terms in the Jacobian.

Neglecting the two terms allows for the following simplification to the characteristic polynomial:

$$\lambda((u-\lambda)^2 c_2 + c_3 + \lambda(u-\lambda)^3 - \lambda(u-\lambda)c_4) = 0 \tag{80}$$

Now, one of the eigenvalues is $\lambda = 0$ and the roots of the remaining fourth-order polynomial can be analytically determined (using, for example, Ferrari's solution). The calculations are extremely laborious and, when comparing the obtained eigenvalues with the numerical solution of (78), some discrepancies can be observed.

Figures 7 and 8 present a comparison between the simplified characteristic polynomial (80) and the complete polynomial (78). The result in Figure 7 is for a generic supersonic flow with $u/a_\infty = 1.61$ and $p/(\rho_\infty a_\infty^2) = 4.51$, while the result in Figure 8 is for a generic subsonic flow with $u/a_\infty = 0.82$ and $p/(\rho_\infty a_\infty^2) = 10.1$. The results indicate that the degree to which the characteristic polynomial roots match is, unfortunately, not constant regardless of the flow scenario considered. Furthermore, even if the characteristic polynomial (78) is to be used, numerically determining the roots of the polynomial via an adequate root finding algorithm is much to numerically expensive to be considered anything close to a viable approach during the evolution stage of the AF method. Therefore, an alternative to using the full Jacobian matrix (75) is required.

While determining an analytical solution for the eigenvalues of the full Jacobian of the hyperbolic Navier-Stokes system still proves elusive, it is possible to check, through numerical calculations, that the eigenvalues are real, at least for typical ranges of variable values. While this does not represent a definitive proof of the hyperbolic nature of the system, it does provide a useful check and a degree of confidence in proceeding with the algorithm.

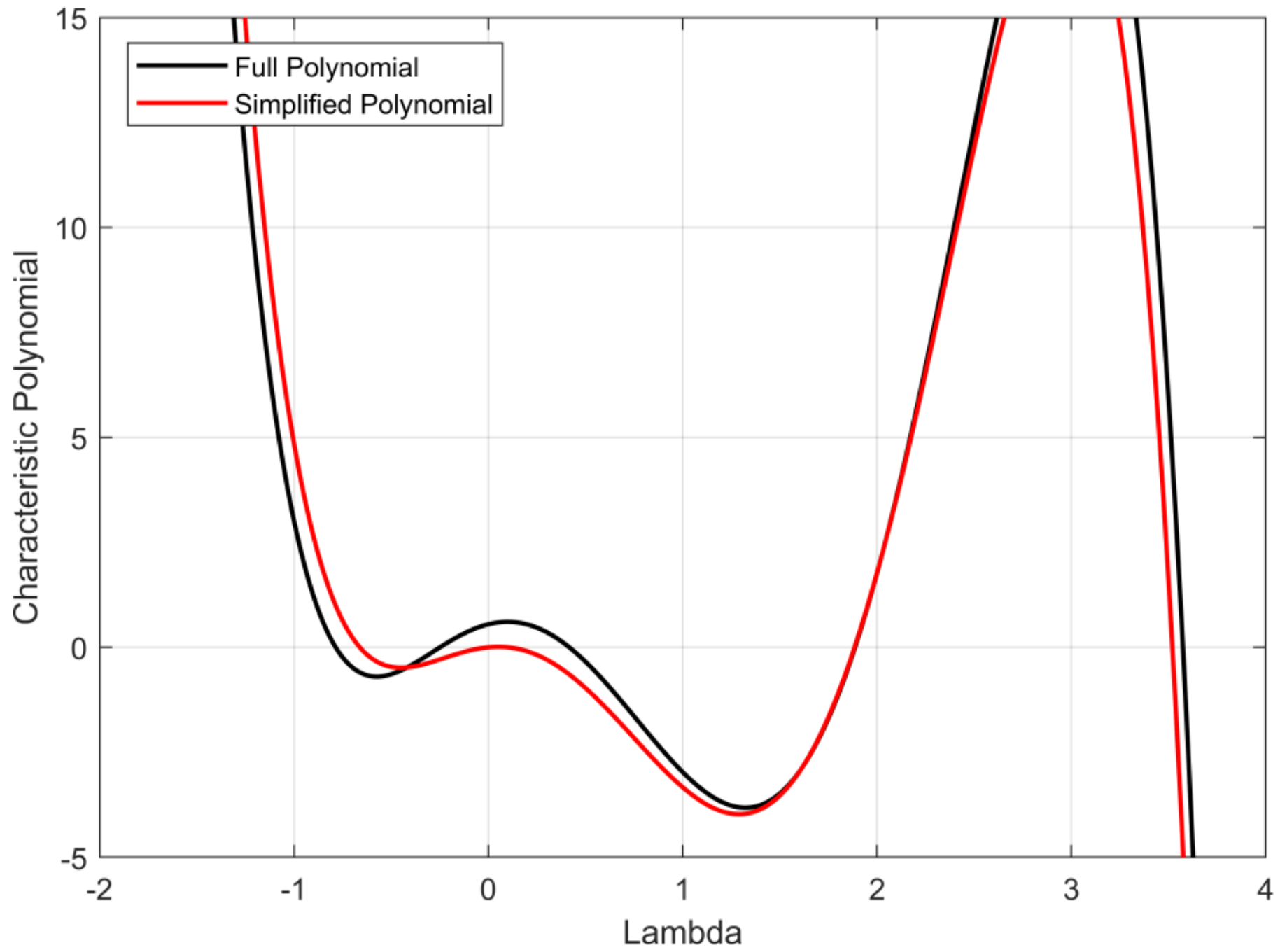

**Figure 7. Complete and simplified characteristic polynomial of the hyperbolic Navier-Stokes equations for a generic supersonic flow.**

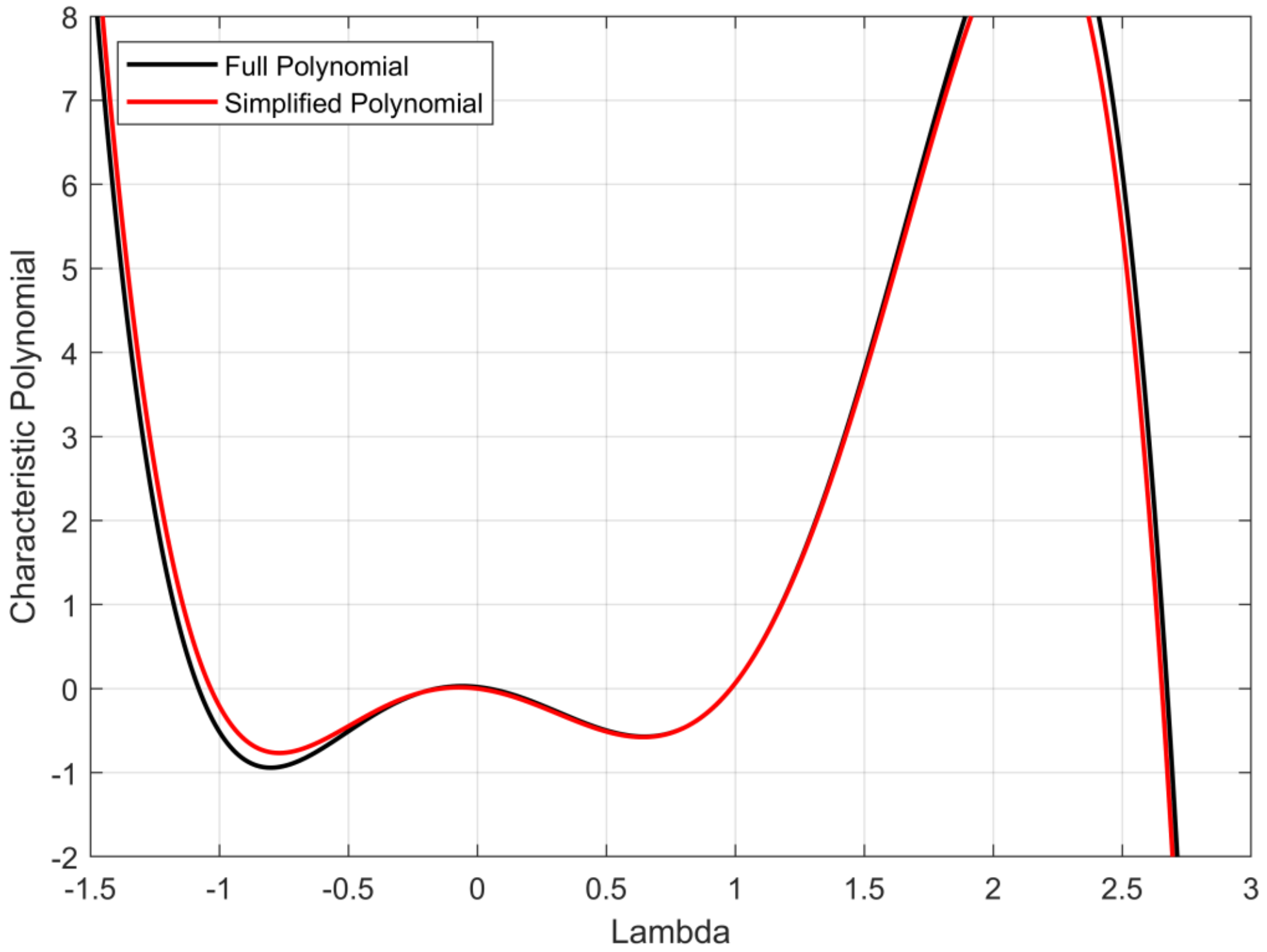

**Figure 8. Complete and simplified characteristic polynomial of the hyperbolic Navier-Stokes equations for a generic subsonic flow.**

The numerical check was done for the following typical ranges of variables values:

$$\begin{gathered}\frac{u}{a_\infty} \in [0, 5]\\ \frac{\mu}{\mu_\infty} \in [0.1, 10]\\ \frac{T}{T_\infty} \in [0.1, 10]\\ \frac{\rho}{\rho_\infty} \in [0.1, 20]\\ \frac{\tau}{\mu_\infty a_\infty / L_\infty} \in [-3, 3]\end{gathered} \tag{81}$$

For each variable, 15 equally spaced values between the upper and lower limits were selected. The coefficients (79) and the characteristic polynomial (78) were generated for all possible combinations between the above-defined variable values, resulting in close to 760,000 distinct cases. For each case, the roots of the polynomial were determined by computing the eigenvalues of the companion matrix, using MATLAB. It was verified that all five roots are real for all cases considered here.

### 5.3. An operator splitting approach

An operator splitting approach is considered as an alternative. Split the flux vector in inviscid and viscous components:

$$\boldsymbol{F} = \boldsymbol{F^I} + \boldsymbol{F^V} = \begin{bmatrix} \rho u \\ \rho u u + p \\ \rho u H \\ 0 \\ 0 \end{bmatrix} + \begin{bmatrix} 0 \\ -\tau \\ -\tau u + q \\ -u \\ \frac{1}{\gamma - 1}\frac{p}{\rho} \end{bmatrix} \tag{82}$$

The hyperbolic formulation of the Navier-Stokes equations is then:

$$\frac{\partial \boldsymbol{U}}{\partial t} + \boldsymbol{P}\frac{\partial \boldsymbol{F^I}}{\partial x} + \boldsymbol{P}\frac{\partial \boldsymbol{F^V}}{\partial x} = \boldsymbol{PS} \tag{83}$$

The split in two distinct problems is written as:

$$\begin{gathered}\frac{\partial \boldsymbol{U}}{\partial t} + \boldsymbol{P}\frac{\partial \boldsymbol{F^I}}{\partial x} = \boldsymbol{0}\\ \frac{\partial \boldsymbol{U}}{\partial t} + \boldsymbol{P}\frac{\partial \boldsymbol{F^V}}{\partial x} = \boldsymbol{PS}\end{gathered} \tag{84}$$

### 5.4. The inviscid flow problem

The inviscid flux $\boldsymbol{F^I}$ is the Euler flux, expanded by two additional zero. The three non-zero eigenvalues and eigenvectors of the inviscid Jacobian $\boldsymbol{P}\partial \boldsymbol{F^I}/\partial \boldsymbol{U}$ are the same as those of the Euler equations. The preconditioning matrix $\boldsymbol{P}$ defined in (74) has no influence on the inviscid eigenvalues and eigenvectors, as the diagonal terms including the relaxation times multiply only null elements of $\boldsymbol{F^I}$.

Evolution of the point degrees of freedom on the cell boundaries is done using the same predictor-corrector scheme presented for the Euler equations:

$$
\begin{aligned}
x_0^{lk} &= x - \frac{\lambda_l + \lambda_k}{2}\Delta t\,, l,k = 1,m \\
\boldsymbol{U}^{(l)}(t^{n+1}, x) &= \sum_{k=1}^{m} \boldsymbol{R^I I}_k \boldsymbol{L^I U}\left(t^n, x_0^{lk}\right)\,, l = 1,m \\
x_0^k &= x - \lambda_k^* \Delta t\,, k = 1,m \\
\boldsymbol{U}(t^{n+1}, x) &= \sum_{k=1}^{m} \boldsymbol{R}_k^{I^*} \boldsymbol{L}_k^{I^*} \boldsymbol{U}\left(t^n, x_0^k\right)
\end{aligned}
\tag{85}
$$

The cell-average values at the new time level are obtained as:

$$
\overline{\boldsymbol{U}}_i^{n+1} - \overline{\boldsymbol{U}}_i^n = -\frac{\Delta t}{\Delta x}\left(\overline{\boldsymbol{F^I}}\left(x_{i+1/2}\right) - \overline{\boldsymbol{F^I}}\left(x_{i-1/2}\right)\right) \tag{86}
$$

Boundary conditions (Dirichlet or Neumann) for the primitive or conservative variables are specified in fictitious ghost cells, having indices $0$ and $N+1$, with the variable values for the first and last faces of the domain, $x_{1/2}$ and $x_{N+1/2}$ being evolved based on the local characteristics. Characteristics-based boundary conditions for the Euler equations are widely used (see Blazek (2015), for example).

### 5.5 The viscous flow problem

The viscous Jacobian is:

$$
\boldsymbol{P}\frac{\partial \boldsymbol{F^V}}{\partial \boldsymbol{U}} = \begin{bmatrix} 0 & 0 & 0 & 0 & 0 \\ 0 & 0 & 0 & -1 & 0 \\ \frac{u\tau}{\rho} & -\frac{\tau}{\rho} & 0 & -u & 1 \\ a_v^2 u & -a_v^2 & 0 & 0 & 0 \\ a_h^2(u^2 - E) & -a_h^2 u & a_h^2 & 0 & 0 \end{bmatrix} \tag{87}
$$

The eigenvalues of the viscous Jacobian are obtained as:

$$
\begin{aligned}
\lambda_{1,2} &= \mp a_v \\
\lambda_{3,4} &= \mp a_h \\
\lambda_5 &= 0
\end{aligned}
\tag{88}
$$

And the corresponding right eigenvectors:

$$
\boldsymbol{R^V} = \begin{bmatrix} 0 & 0 & 0 & 0 & 1 \\ \rho & \rho & 0 & 0 & u \\ \rho u + \frac{\tau Pr_n}{a_v(Pr_n - 1)} & \rho u - \frac{\tau Pr_n}{a_v(Pr_n - 1)} & \frac{1}{\gamma - 1} & \frac{1}{\gamma - 1} & E \\ \rho a_v & -\rho a_v & 0 & 0 & 0 \\ -\frac{\tau}{Pr_n - 1} & -\frac{\tau}{Pr_n - 1} & -\frac{a_h}{\gamma - 1} & \frac{a_h}{\gamma - 1} & 0 \end{bmatrix} \tag{89}
$$

where $Pr_n = a_v^2/a_h^2$.

The left eigenvectors are:

$$L^{V} = \begin{bmatrix} -\frac{u}{2\rho} & \frac{1}{2\rho} & 0 & \frac{1}{2\rho a_v} & 0 \\ -\frac{u}{2\rho} & \frac{1}{2\rho} & 0 & -\frac{1}{2\rho a_v} & 0 \\ \left[\frac{u\tau}{Pr_n - 1} + \rho a_h \left(u^2 - E\right)\right]\frac{\gamma - 1}{2\rho a_h} & -\left(\frac{\tau}{Pr_n - 1} + \rho a_h u\right)\frac{\gamma - 1}{2\rho a_h} & \frac{\gamma - 1}{2} & -\frac{\tau Pr_n}{a_v\left(Pr_n - 1\right)}\frac{\gamma - 1}{2\rho a_v} & -\frac{\gamma - 1}{2a_h} \\ -\left[\frac{u\tau}{Pr_n - 1} - \rho a_h \left(u^2 - E\right)\right]\frac{\gamma - 1}{2\rho a_h} & \left(\frac{\tau}{Pr_n - 1} - \rho a_h u\right)\frac{\gamma - 1}{2\rho a_h} & \frac{\gamma - 1}{2} & -\frac{\tau Pr_n}{a_v\left(Pr_n - 1\right)}\frac{\gamma - 1}{2\rho a_v} & \frac{\gamma - 1}{2a_h} \\ 1 & 0 & 0 & 0 & 0 \end{bmatrix} \qquad (90)$$

The following notations are introduced:

$$\tau_n = \frac{\tau}{Pr_n - 1} \quad b_h = \frac{\gamma - 1}{2\rho a_h} \quad b_v = \frac{\gamma - 1}{2\rho a_v} \quad a_p = \frac{\tau Pr_n}{a_v(Pr_n - 1)}$$
$$c^+ = u\tau_n + \rho a_h\left(\frac{u^2}{2} - \frac{a^2}{\gamma(\gamma-1)}\right) \quad c^- = u\tau_n - \rho a_h\left(\frac{u^2}{2} - \frac{a^2}{\gamma(\gamma-1)}\right) \tag{91}$$
$$\tau^+ = \tau_n + \rho a_h u \quad \tau^- = \tau_n - \rho a_h u$$

After some calculations, the characteristic variables for the viscous subproblem can be written as follows:

$$\partial \boldsymbol{W^V} = \begin{bmatrix} -\frac{u\partial\rho}{2\rho} + \frac{\partial u}{2\rho} + \frac{b_v\partial\tau}{\gamma - 1} \\ -\frac{u\partial\rho}{2\rho} + \frac{\partial u}{2\rho} - \frac{b_v\partial\tau}{\gamma - 1} \\ c^+b_h\partial\rho - \tau^+b_h\partial u + \frac{\gamma-1}{2}\partial p - a_pb_v\partial\tau - \rho b_v\partial q \\ -c^-b_h\partial\rho + \tau^-b_h\partial u + \frac{\gamma-1}{2}\partial p - a_pb_v\partial\tau + \rho b_v\partial q \\ \partial\rho \end{bmatrix} \tag{92}$$

Evolution of the point degrees of freedom on the cell boundaries is done using the same predictor-corrector scheme presented for the hyperbolic diffusion equation:

$$\begin{aligned}
x_0^{lk} &= x - \frac{\lambda_l + \lambda_k}{2}\Delta t\,, l,k = 1,m \\
x2_0^{lk} &= x - \frac{\lambda_l + \lambda_k}{2}\frac{\Delta t}{2}\,, l,k = 1,m \\
\boldsymbol{U}^{(l)}(t^{n+1},x) &= \sum_{k=1}^{m}\Big[\boldsymbol{R^V I}_k\boldsymbol{L^V U}\big(t^n,x_0^{lk}\big) + \frac{\lambda_k\Delta t}{6}\boldsymbol{R^V I}_k\boldsymbol{L^V Q}\big(t^n,x_0^{lk}\big) \\
&\quad + \frac{2\lambda_k\Delta t}{3}\boldsymbol{R^V I}_k\boldsymbol{L^V Q}\big(t^n,x2_0^{lk}\big) + \frac{\lambda_k\Delta t}{6}\boldsymbol{R^V I}_k\boldsymbol{L^V Q}(t^n,x)\Big]\,, l = 1,m \\
x_0^k &= x - \lambda_k^*\Delta t\,, k = 1,m \\
x2_0^k &= x - \lambda_k^*\frac{\Delta t}{2}\,, k = 1,m \\
\boldsymbol{U}(t^{n+1},x) &= \Big[\sum_{k=1}^{m}\boldsymbol{R}_k^{\boldsymbol{V}^*}\boldsymbol{L}_k^{\boldsymbol{V}^*}\boldsymbol{U}\big(t^n,x_0^k\big) + \frac{\lambda_k^*\Delta t}{6}\boldsymbol{R}_k^{\boldsymbol{V}^*}\boldsymbol{L}_k^{\boldsymbol{V}^*}\boldsymbol{Q}\big(t^n,x_0^k\big) \\
&\quad + \frac{2\lambda_k^*\Delta t}{3}\boldsymbol{R}_k^{\boldsymbol{V}^*}\boldsymbol{L}_k^{\boldsymbol{V}^*}\boldsymbol{Q}\big(t^n,x2_0^k\big) + \frac{\lambda_k^*\Delta t}{6}\boldsymbol{R}_k^{\boldsymbol{V}^*}\boldsymbol{L}_k^{\boldsymbol{V}^*}\boldsymbol{Q}(t^n,x)\Big]
\end{aligned} \tag{93}$$

The cell-average values at the new time level are obtained by converging the dual time problem to steady state over one physical time step:

$$\begin{aligned}
\overline{\boldsymbol{U}}_i^{k+1} - \overline{\boldsymbol{U}}_i^k &= -\frac{\alpha\Delta\tau}{\Delta t}\overline{\boldsymbol{U}}_i^{k+1} + \Delta\tau\boldsymbol{R}^* \\
\boldsymbol{R}^* &= -\frac{1}{\Delta x}\Big(\overline{\boldsymbol{F^V}}\big(x_{i+1/2}\big) - \overline{\boldsymbol{F^V}}\big(x_{i-1/2}\big)\Big) + \frac{1}{\Delta t\Delta x}\int_{t^n}^{t^{n+1}}\int_{x_{i-1/2}}^{x_{i+1/2}}\boldsymbol{S}dx\,dt - \frac{\beta}{\Delta t}\overline{\boldsymbol{U}}_i^n \\
&\quad - \frac{\gamma}{\Delta t}\overline{\boldsymbol{U}}_i^{n-1} - \frac{\delta}{\Delta t}\overline{\boldsymbol{U}}_i^{n-2}
\end{aligned} \tag{94}$$

$$\bar{\boldsymbol{U}}_i^{n+1} = \bar{\boldsymbol{U}}_i^{k+1}$$

To implement the boundary conditions, it will first be noted that the first and third eigenvalues are always negative, while the second and fourth eigenvalues are always positive.

The characteristic curves associated with the second and fourth eigenvalues will enter the domain at the left boundary $x_{1/2}$. Since these characteristic curves bring information from outside the domain (from the ghost cell), the effect of the source term should not be included, giving:

$$\begin{aligned} \partial W_2(t^{n+1}, x_{1/2}) &= \partial W_2(t^n, x_{1/2} - \lambda_2 \Delta t) \\ \partial W_4(t^{n+1}, x_{1/2}) &= \partial W_4(t^n, x_{1/2} - \lambda_4 \Delta t) \end{aligned} \tag{95}$$

The first and third characteristic curve brings information from inside the domain, so the effect of the source term must be included:

$$\begin{aligned} \partial W_1(t^{n+1}, x_{1/2}) &= \partial W_1(t^n, x_{1/2} - \lambda_1 \Delta t) + \int_{x_{1/2}-\lambda_1\Delta t}^{x_{1/2}} \frac{S_1^w(t^n, x)}{\lambda_1} dx \\ \partial W_3(t^{n+1}, x_{1/2}) &= \partial W_3(t^n, x_{1/2} - \lambda_3 \Delta t) + \int_{x_{1/2}-\lambda_3\Delta t}^{x_{1/2}} \frac{S_3^w(t^n, x)}{\lambda_3} dx \end{aligned} \tag{96}$$

The integral can be determined as presented previously, using data from the first cell of the domain. For ease of notation, let us denote:

$$\int_{x_{1/2}-\lambda_1\Delta t}^{x_{1/2}} \frac{S_1^w(t^n, x)}{\lambda_1} dx = I_1 \qquad \int_{x_{1/2}-\lambda_3\Delta t}^{x_{1/2}} \frac{S_3^w(t^n, x)}{\lambda_3} dx = I_3 \tag{97}$$

Based on the fifth characteristic variable, the density values on the boundary $\rho_{1/2}$ are considered known and taken from the inviscid flow problem boundary conditions. Following the approach of Whitfield and Janus (1984) for deriving characteristics-based boundary conditions, and using a subscript $0$ to denote values from the ghost cell and subscript $d$ to denote values from inside the first cell, the following set of boundary conditions is obtained:

$$\begin{aligned} u_{1/2} &= \frac{1}{2}(u_d + u_0) + \frac{u}{2}(2\,\rho_{1/2} - \rho_0 - \rho_d) + \frac{\rho b_v}{\gamma - 1}(\tau_d - \tau_0) + \rho I_1 \\ \tau_{1/2} &= \tau_0 + \frac{(\gamma - 1)u}{2\rho b_v}(\rho_0 - \rho_{1/2}) + \frac{\gamma - 1}{2\rho b_v}(u_{1/2} - u_0) \\ q_{1/2} &= \frac{1}{2}(q_0 + q_d) + \frac{c^+ b_h}{2\rho b_v}(\rho_{1/2} - \rho_d) - \frac{c^- b_h}{2\rho b_v}\rho_0 + \frac{\gamma - 1}{4\rho b_v}(p_0 - p_d) \\ &\quad - \frac{\tau^+ b_h}{2\rho b_v}(u_{1/2} - u_d) - \frac{\tau^- b_h}{2\rho b_v}(u_{1/2} - u_0) + \frac{a_p}{2\rho}(\tau_d - \tau_0) - \frac{I_3}{2\rho b_v} \\ p_{1/2} &= p_0 + \frac{2c^- b_h}{\gamma - 1}(\rho_{1/2} - \rho_0) - \frac{2\tau^- b_h}{\gamma - 1}(u_{1/2} - u_0) + \frac{2a_p b_v}{\gamma - 1}(\tau_{1/2} - \tau_0) \end{aligned} \tag{98}$$

The various terms such as $u/2$, $\rho b_v/(\gamma - 1)$, $c^+ b_h/(2\rho b_v\,)$ and others are calculated using a reference state, normally set using values from an interior point (such as point $d$). A similar approach is used for right boundary of the domain, $x_{N+1/2}$.

## 5.6. Achieving third order accurate solution after operator splitting

It must be ensured that the operator splitting approach has the desired order of accuracy. If only first or second-order accuracy in time are desired, the classical splitting schemes such as Lie or Strang splitting can be used. It was shown by Jia and Li (2011) that higher order accuracy results can be obtained via operator splitting by taking a suitable combination of first order and second-order results.

Consider the example ordinary differential equation:

$$\frac{dU}{dt} = F_1(U) + F_2(U) \tag{99}$$

One of the best know splitting techniques is the Lie splitting (very well explained by Trotter, 1959), which is first-order accurate in time and consists of the following steps:

$$\begin{gathered}
assume\ U(t^n)\ is\ known \\
solve\ \frac{dU^*}{dt} = F_1(U^*)\ between\ t^n\ and\ t^{n+1} \\
with\ initial\ condition\ U^*(t^n) = U(t^n) \\
solve\ \frac{dU^{**}}{dt} = F_2(U^{**})\ between\ t^n\ and\ t^{n+1} \\
with\ initial\ condition\ U^{**}(t^n) = U^*(t^{n+1}) \\
set\ U(t^{n+1}) = U^{**}(t^{n+1})
\end{gathered} \tag{100}$$

An equally well-known splitting technique, which is second-order accurate in time, is the Strang splitting (Strang, 1968):

$$\begin{gathered}
assume\ U(t^n)\ is\ known \\
solve\ \frac{dU^*}{dt} = F_1(U^*)\ between\ t^n\ and\ t^{n+1/2} \\
with\ initial\ condition\ U^*(t^n) = U(t^n) \\
solve\ \frac{dU^{**}}{dt} = F_2(U^{**})\ between\ t^n\ and\ t^{n+1} \\
with\ initial\ condition\ U^{**}(t^n) = U^*\left(t^{n+1/2}\right) \\
solve\ \frac{dU^{***}}{dt} = F_1(U^{***})\ between\ t^{n+1/2}\ and\ t^{n+1} \\
with\ initial\ condition\ U^{***}\left(t^{n+1/2}\right) = U^{**}(t^{n+1}) \\
set\ U(t^{n+1}) = U^{***}(t^{n+1})
\end{gathered} \tag{101}$$

Consider the following combination of splitting techniques:

$$\begin{gathered}
solve\ using\ Lie\ splitting\ and\ the\ order\ F_1 \to F_2\ to\ get\ U_1 \\
solve\ using\ Lie\ splitting\ and\ the\ order\ F_2 \to F_1\ to\ get\ U_2 \\
solve\ using\ Strang\ splitting\ and\ the\ order\ F_1 \to F_2 \to F_1\ to\ get\ U_3 \\
solve\ using\ Strang\ splitting\ and\ the\ order\ F_2 \to F_1 \to F_2\ to\ get\ U_4
\end{gathered} \tag{102}$$

Finally,

$$U(t^{n+1}) = \frac{2}{3}(U_3 + U_4) - \frac{1}{6}(U_1 + U_2) \tag{103}$$

To show that the solution in (103) is third-order accurate, the approach proposed by Jia and Li (2011) is used.

First, (99) is written as two operators acting on the variable of interest:

$$\frac{dU}{dt} = (F_1 + F_2) * U \tag{104}$$

The solution over one time step is given by:

$$U(t^{n+1}) = e^{\Delta t(F_1+F_2)} * U(t^n) \tag{105}$$

Let $\widehat{U}(t^{n+1})$ be the solution obtained by a splitting approach. The error is then given by:

$$E(t^{n+1}) = U(t^{n+1}) - \widehat{U}(t^{n+1}) = E_{ops} * U(t^n) \tag{106}$$

Here, the newly defined operator $E_{ops}$ is called the splitting error operator. If $E_{ops}$ is of the order of $\mathcal{O}(\Delta t^{p+1})$ the splitting is of $p$ order of accuracy.

The solutions obtained by the four splitting techniques defined in (102) are:

$$\begin{aligned}
\widehat{U}_1(t^{n+1}) &= e^{\Delta t F_2} e^{\Delta t F_1} * U(t^n)\\
\widehat{U}_2(t^{n+1}) &= e^{\Delta t F_1} e^{\Delta t F_2} * U(t^n)\\
\widehat{U}_3(t^{n+1}) &= e^{\frac{\Delta t}{2}F_1} e^{\Delta t F_2} e^{\frac{\Delta t}{2}F_1} * U(t^n)\\
\widehat{U}_4(t^{n+1}) &= e^{\frac{\Delta t}{2}F_2} e^{\Delta t F_1} e^{\frac{\Delta t}{2}F_2} * U(t^n)
\end{aligned} \tag{107}$$

Consider a final solution given by:

$$\widehat{U}(t^{n+1}) = \alpha\big(\widehat{U}_3 + \widehat{U}_4\big) + \beta\big(\widehat{U}_1 + \widehat{U}_2\big) \tag{108}$$

The corresponding splitting error operator is:

$$\begin{aligned}
E_{ops} = e^{\Delta t(F_1+F_2)} - \alpha\left(e^{\frac{\Delta t}{2}F_1} e^{\Delta t F_2} e^{\frac{\Delta t}{2}F_1} + e^{\frac{\Delta t}{2}F_2} e^{\Delta t F_1} e^{\frac{\Delta t}{2}F_2}\right)\\
-\beta\left(e^{\Delta t F_2} e^{\Delta t F_1} + e^{\Delta t F_1} e^{\Delta t F_2}\right)
\end{aligned} \tag{109}$$

Next, the following identities are written using the Baker–Campbell–Hausdorff formula:

$$\begin{aligned}
e^{\Delta t F_1} e^{\Delta t F_2} &= e^{\Delta t F_1 + \Delta t F_2 + \frac{\Delta t^2}{2}[F_1,F_2] + \frac{\Delta t^3}{12}[F_1,F_1,F_2] - \frac{\Delta t^3}{12}[F_2,F_1,F_2] + \mathcal{O}(\Delta t^4)}\\
e^{\Delta t F_2} e^{\Delta t F_1} &= e^{\Delta t F_1 + \Delta t F_2 + \frac{\Delta t^2}{2}[F_2,F_1] + \frac{\Delta t^3}{12}[F_2,F_2,F_1] - \frac{\Delta t^3}{12}[F_1,F_2,F_1] + \mathcal{O}(\Delta t^4)}\\
e^{\frac{\Delta t}{2}F_1} e^{\Delta t F_2} e^{\frac{\Delta t}{2}F_1} &= e^{\Delta t F_1 + \Delta t F_2 + \frac{\Delta t^3}{12}[F_2,F_2,F_1] - \frac{\Delta t^3}{24}[F_1,F_1,F_2] + \mathcal{O}(\Delta t^4)}\\
e^{\frac{\Delta t}{2}F_2} e^{\Delta t F_1} e^{\frac{\Delta t}{2}F_2} &= e^{\Delta t F_1 + \Delta t F_2 + \frac{\Delta t^3}{12}[F_1,F_1,F_2] - \frac{\Delta t^3}{24}[F_2,F_2,F_1] + \mathcal{O}(\Delta t^4)}
\end{aligned} \tag{110}$$

Here, $[F_1, F_2] = F_1F_2 - F_2F_1$ is the commutator, and $[F_1, F_1, F_2] = \big[F_1, [F_1, F_2]\big]$ is a recursive application of the commutator.

A Taylor series expansion is made for each term in (109), and only terms up to $\mathcal{O}(\Delta t^4)$ are kept.

$$
\begin{aligned}
E_{ops} = I + \Delta t(F_1 + F_2) + \frac{\Delta t^2}{2}(F_1 + F_2)^2 + \frac{\Delta t^3}{6}(F_1 + F_2)^3 \\
-2\alpha I - 2\alpha\Delta t(F_1 + F_2) - \alpha\Delta t^2(F_1 + F_2)^2 \\
-\alpha\frac{\Delta t^3}{3}(F_1 + F_2)^3 - \alpha\frac{\Delta t^3}{24}\{[F_2, F_2, F_1] + [F_1, F_1, F_2]\} \\
-2\beta I - 2\beta\Delta t(F_1 + F_2) - \beta\Delta t^2(F_1 + F_2)^2 - \beta\frac{\Delta t^2}{2}\{[F_1, F_2] + [F_2, F_1]\} \\
-\beta\frac{\Delta t^3}{3}(F_1 + F_2)^3 \\
-\beta\frac{\Delta t^3}{12}\{[F_1, F_1, F_2] + [F_2, F_2, F_1] - [F_2, F_1, F_2] - [F_1, F_2, F_1]\} \\
-\beta\frac{\Delta t^3}{4}\{(F_1 + F_2)[F_1, F_2] + [F_1, F_2](F_1 + F_2)\} \\
-\beta\frac{\Delta t^3}{4}\{(F_1 + F_2)[F_2, F_1] + [F_2, F_1](F_1 + F_2)\} \\
+\mathcal{O}(\Delta t^4)
\end{aligned} \tag{111}
$$

This is also written as:

$$
\begin{aligned}
E_{ops} = (1 - 2\alpha - 2\beta)I + (1 - 2\alpha - 2\beta)\Delta t(F_1 + F_2) \\
+\Delta t^2\left\{\frac{1}{2}(F_1 + F_2)^2 - \alpha(F_1 + F_2)^2 - \beta(F_1 + F_2)^2 - \frac{\beta}{2}\{[F_1, F_2] + [F_2, F_1]\}\right\} \\
+\Delta t^3\left\{\frac{1}{6}(F_1 + F_2)^3 - \frac{\alpha}{3}(F_1 + F_2)^3 - \frac{\alpha}{24}\{[F_2, F_2, F_1] + [F_1, F_1, F_2]\}\right. \\
-\frac{\beta}{3}(F_1 + F_2)^3 \\
-\frac{\beta}{12}\{[F_1, F_1, F_2] + [F_2, F_2, F_1] - [F_2, F_1, F_2] - [F_1, F_2, F_1]\} \\
-\frac{\beta}{4}\{(F_1 + F_2)[F_1, F_2] + [F_1, F_2](F_1 + F_2)\} \\
\left.-\frac{\beta}{4}\{(F_1 + F_2)[F_2, F_1] + [F_2, F_1](F_1 + F_2)\}\right\} \\
+\mathcal{O}(\Delta t^4)
\end{aligned} \tag{112}
$$

To achieve third-order accuracy, all terms present in (112) up to and including $\Delta t^3$ terms must cancel out. It is relatively easy to check that all combinations of $\alpha$ and $\beta$ that satisfy the condition $2\alpha + 2\beta = 1$ achieve second-order accuracy. However, calculations show that only the choice $\alpha = 2/3$ and $\beta = -1/6$ make the third-order terms cancel out. The calculations are extremely long and are not included here.

Hence, $E_{ops} = \mathcal{O}(\Delta t^4)$ for the combination of splitting techniques (102)-(103) and the obtained solution is third-order accurate.

### 5.7. Exact solution test problem

The performance of the AF scheme for the Navier-Stokes equations is investigated using a manufactured exact solution.

Consider the following functions:

$$
\begin{aligned}
\rho &= 1 + 0.1t\sin x \\
u &= 1 + 2\sin x \\
p &= 1 - 0.5\cos x
\end{aligned}
\tag{113}
$$

The solution is defined in a domain $x \in [-\pi, \pi]$. Total energy is calculated based on the above primitive variables using $\rho E = p/(\gamma - 1) + \rho u^2/2$.

The functions defined above are then inserted in (69) and the following source terms are found:

$$
\begin{aligned}
S_1 &= 0.1\sin x + (2 + 0.1t + 0.4t\sin x)\cos x \\
S_2 &= \left(0.6 + \frac{8}{3}\mu + 0.2\sin x\right)\sin x \\
&\quad + (4 + 0.1t + 8\sin x + 0.8t\sin x + 1.2t\sin^2 x)\cos x \\
S_3 &= \left(0.05 + 0.2\sin x + 0.2\sin^2 x + \frac{0.5\gamma}{\gamma - 1} + \frac{8}{3}\mu\right)\sin x \\
&\quad + \left(\frac{5\gamma - 3}{\gamma - 1} + 0.05t + 0.6t\sin x + 12\sin x + 1.8t\sin^2 x + 12\sin^2 x \right. \\
&\quad \left. + 1.6t\sin^3 x\right)\cos x \\
&\quad - \left(\frac{\gamma}{\gamma - 1} + \frac{16}{3}\mu\right)(\cos^2 x - \sin^2 x) \\
&\quad - \frac{\gamma\mu}{Pr(\gamma - 1)}\left[\frac{(0.5\cos x + 0.1t\sin x)(1 + 0.1t\sin x)}{(1 + 0.1t\sin x)^3} \right. \\
&\quad \left. - \frac{0.2t\cos x\,(0.05t + 0.5\sin x - 0.1t\cos x)}{(1 + 0.1t\sin x)^3}\right]
\end{aligned}
\tag{114}
$$

The final source term of the hyperbolic formulation is then set as indicated in (115) below. Solving the Navier-Stokes equations with this specific source term will ensure an exact solution is obtained as defined in (113).

$$
\boldsymbol{S} = \begin{bmatrix} S_1 \\ S_2 \\ S_3 \\ -\dfrac{\tau}{\mu_V} \\ -\dfrac{q}{\mu_H} \end{bmatrix}
\tag{115}
$$

The obtained solution has no relevant physical significance and is only used for the purpose of verifying the order of convergence for a continuous solution. The $L2$ norm of the density error between the solution at increasing grid density and the exact solution is calculated at $t = 0.8$ and plotted in Figure 9. The coarsest grid level has $N = 5$ cells and the number of cells is doubled for each subsequent grid level. It can be observed that the proposed AF scheme achieves convergence at third-order accuracy for the Navier-Stokes manufactured solution.

### 5.8. Shu-Osher test problem

Another test case considered is a viscous variation of the challenging Shu-Osher shock tube problem (Shu and Osher, 1989). Consider $x \in [-5\,,5]$ and the following (nondimensional) initial conditions:

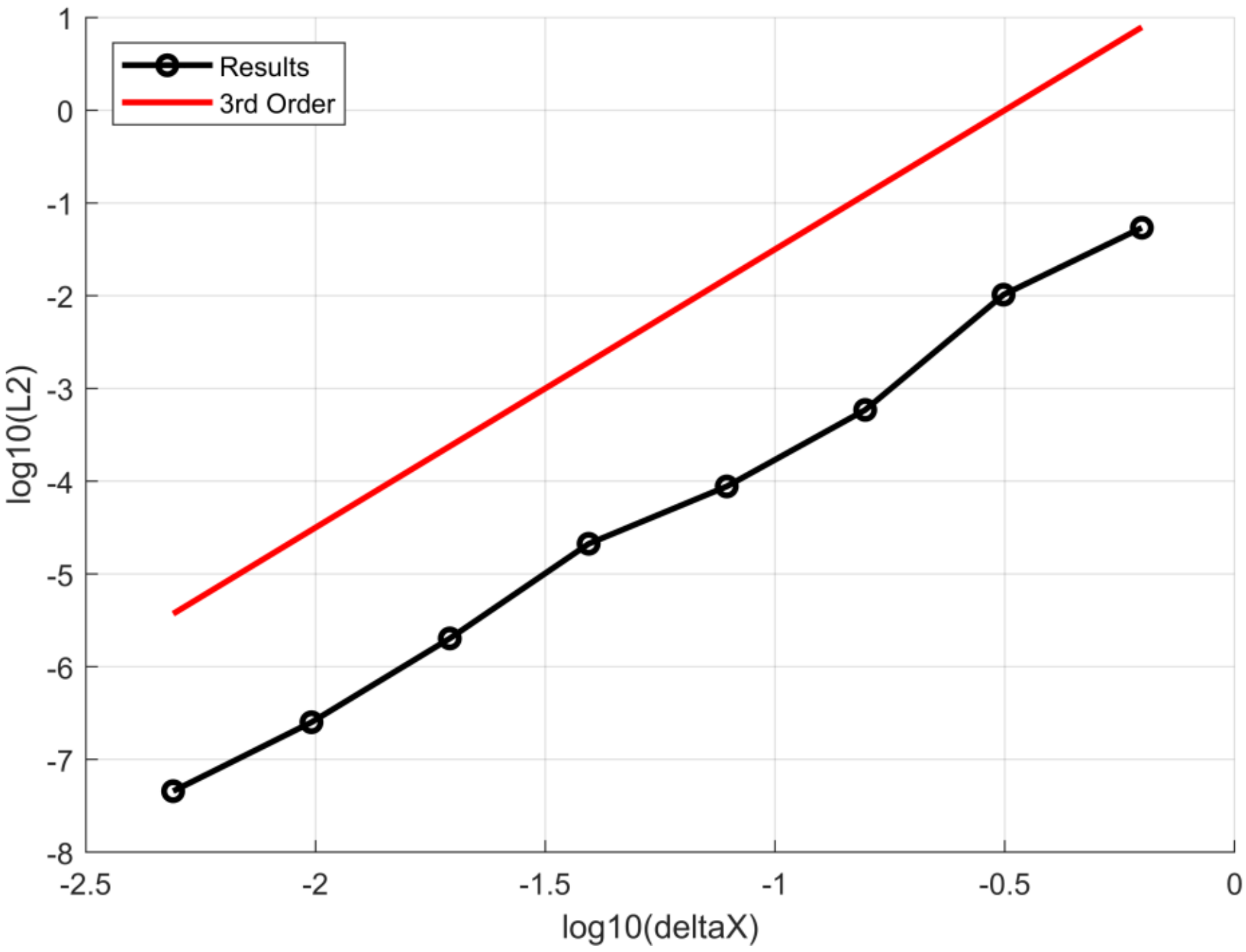

**Figure 9. Order of convergence of the unsteady solution of the hyperbolic Navier-Stokes system measured by the $L2$ error with respect to the manufactured exact solution. A line corresponding to 3rd order is shown for reference.**

$$\begin{cases} \rho = 3.857143\,, p = 10.33333\,, u = 2.629369\; for\; x \leq -4 \\ \rho = 1 + 0.2\sin(\pi x)\;, p = 1\,, u = 0\; for\; x > -4 \end{cases} \quad (116)$$

The numerical simulation is done using $N = 20$ and $N = 100$ grid cells, a Courant number $\upsilon = 0.5$ for a (nondimensional) maximum time of $t = 1.8$. Figure 10 presents a comparison between the AF scheme results using the above-mentioned parameters and a reference simulation using $N = 600$ grid cells. This approach of comparing a coarser grid solution with a finer grid solution is typical for the Shu-Osher problem, as no analytical solution exists. It is observed that the AF scheme can capture the interaction between the shock wave and the sinusoidal pressure wave with relatively few degrees of freedom. The $N = 20$ grid is insufficient to accurately capture the oscillations behind the shock, but there are relatively few differences between the $N = 100$ and $N = 600$ grid. Typical discontinuous FV schemes and require finer grid resolution to predict this test case (see, for example, Cockburn et al., 1989). Again, it must be remembered that the results shown in Figure 10 were obtained without any limiting or dissipation, highlighting the AF scheme's favourable behaviour for problems involving discontinuous solutions.

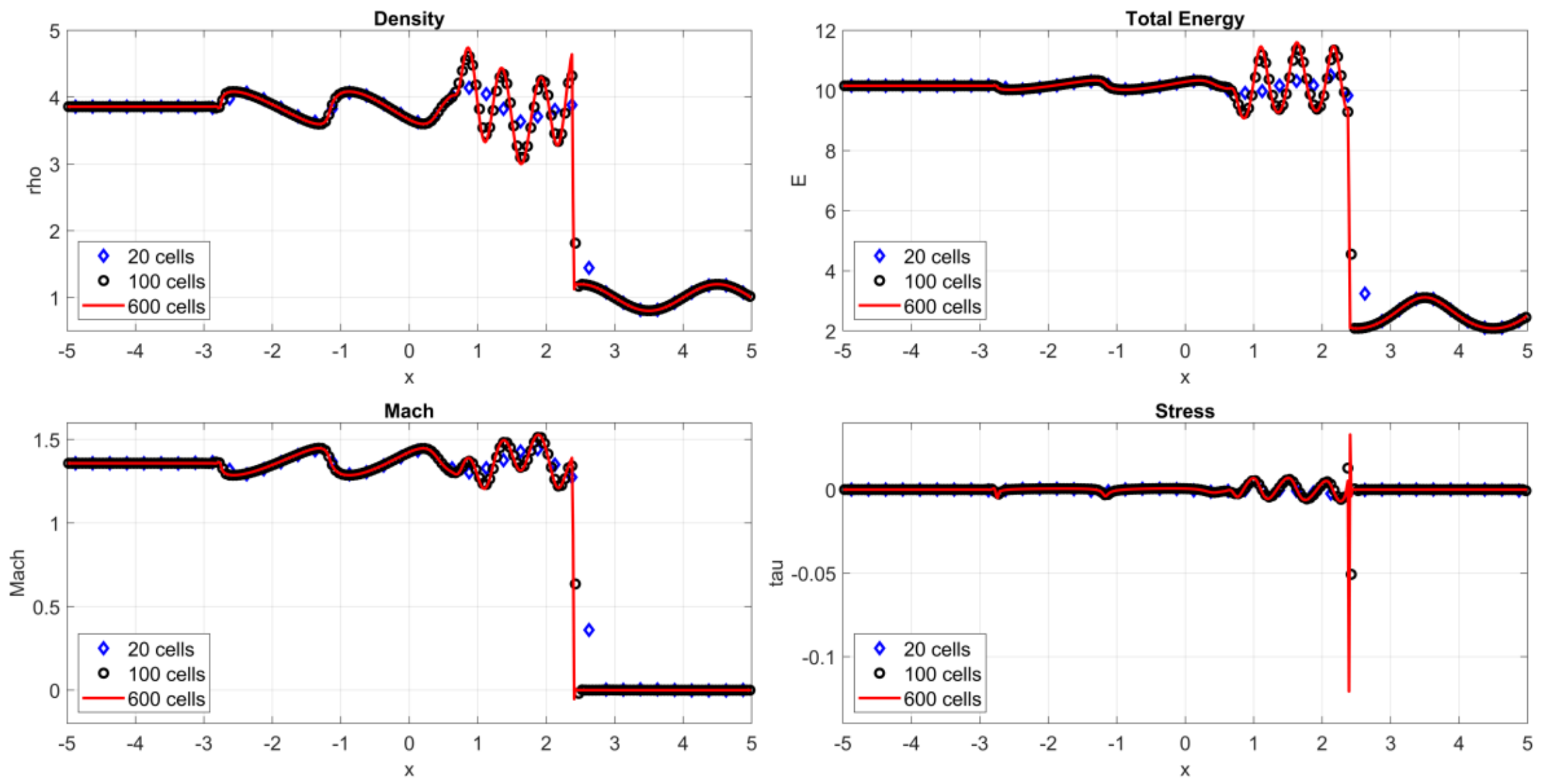

**Figure 10. Shu-Osher shock tube problem for the Navier-Stokes equations. Solution was obtained at $t = 1.8$ with $N = 20$, $N = 100$ and $N = 600$ grid cells and $\nu = 0.5$. Plotted results include both boundary point values and cell-average values.**

Finally, the order of convergence is checked. The $L2$ norm of the density error between the solution at increasing grid density and the solution calculated on a grid with $N = 2560$ cells at $t = 1.8$ is calculated and plotted in Figure 11. The coarsest grid level has $N = 10$ cells and the number of cells is doubled for each subsequent grid level. Just like the application on the Euler equations, the order of accuracy of the scheme is reduced and the AF scheme converges at first-order accuracy for this test case involving moving discontinuities.

## 6. CONCLUSIONS AND FUTURE WORK

The Active Flux scheme is a Finite Volume scheme with additional degrees of freedom. It makes use of a continuous reconstruction and does not require a Riemann solver. An evolution operator is used for the additional degrees of freedom on the cell boundaries. For nonlinear equations and systems, approximate evolution operators must be constructed. It was seen that an efficient evolution operator can be defined for both linear and nonlinear hyperbolic systems of equations and was extended to include the effects of linear source terms. The Active Flux scheme is third-order accurate in space, as verified by the results obtained, though it is possible to extend to even higher orders of accuracy. For the numerical computation of compressible viscous flows, an operator splitting approach achieving third order of accuracy can be derived. Overall, the Active Flux scheme shows a very good ability to capture complex flow physics with good accuracy and fewer degrees of freedom compared to other high-order schemes.

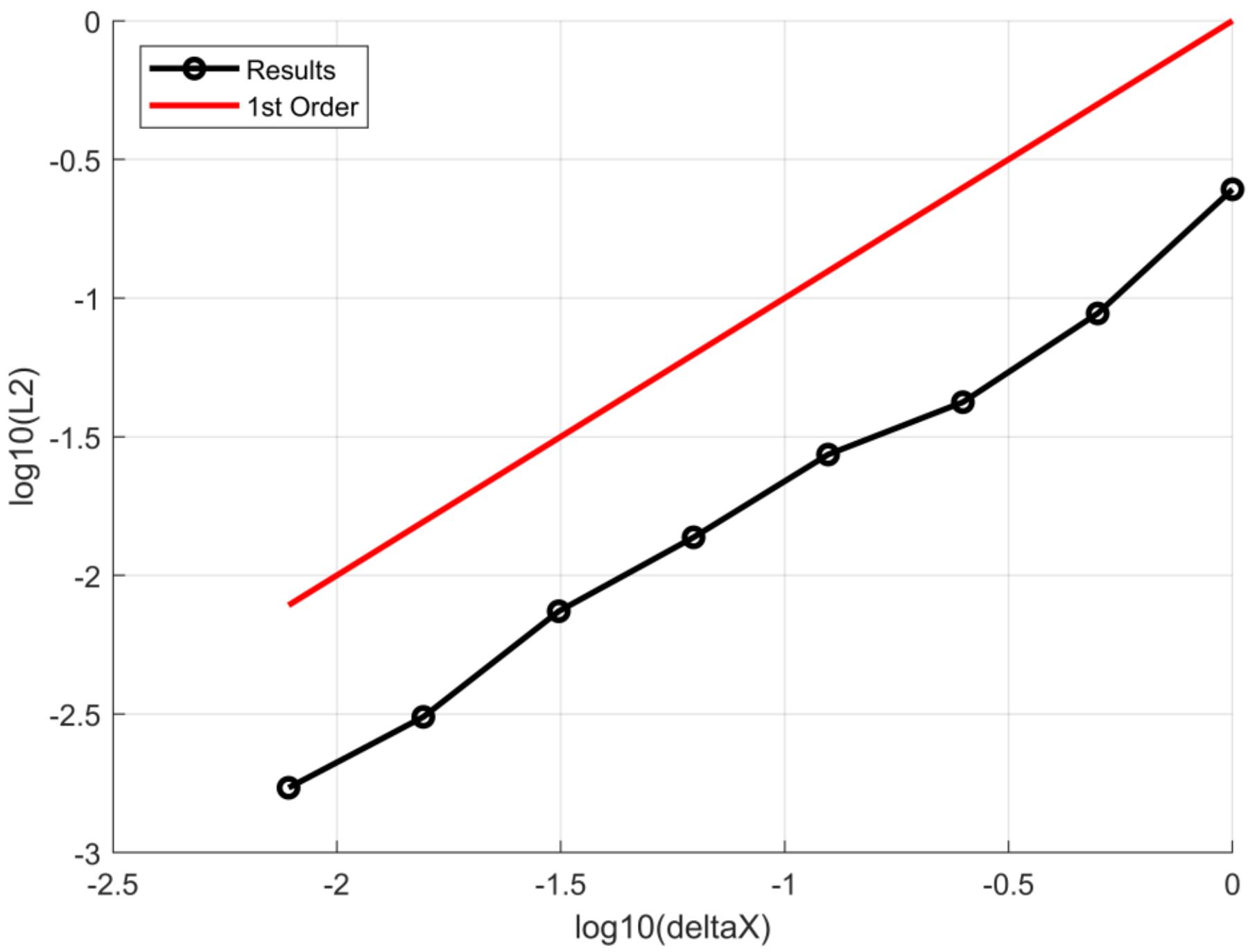

**Figure 11. Order of convergence of the solution to Navier-Stokes equations at $t = 1.8$ as measured by the $L2$ error with respect to the solution on a grid having $N = 2560$ cells. A line corresponding to 1$^{st}$ order is shown for reference.**

Numerous directions exist for future work. Finding a suitable limiter that follows the philosophy of the Active Flux scheme (simple, compact, physically driven) is one such direction. Utilising an implicit dual-time method for time accurate computations is another. Further work needs to be done on determining the eigenvalues and eigenvectors of the full Jacobian matrix of the hyperbolic Navier-Stokes formulation. While the operator splitting approach achieves the desired order of accuracy, it is not computationally efficient due to the need to obtain four distinct solutions at each time step. This can only be alleviated by working with the eigenvectors of the full Jacobian matrix. Finally, extensions of the scheme (including the evolution operator and using characteristic cones instead of characteristic lines) to multi-dimensional flows need to be investigated.

## DECLARATIONS

### Funding and/or Competing Interests

No funding was received to assist with the preparation of this manuscript.

The author has no competing interests to declare that are relevant to the content of this article.

### Data Availability

Data can be made available from the corresponding author on reasonable request.